\documentclass[12pt,english]{article}

\usepackage[hyphens]{url}             
\usepackage{afterpage}
\usepackage{float}
\usepackage{etoolbox}
\usepackage[skip=0.1\baselineskip]{caption}

\usepackage[letterpaper,margin=1.25in]{geometry}

\usepackage{placeins}
\usepackage{subcaption}
\usepackage[T1]{fontenc}
\usepackage[utf8]{inputenc}
\usepackage{babel}
\usepackage{array}
\usepackage{booktabs}
\usepackage{units}
\usepackage{amsmath}
\usepackage{amssymb}
\usepackage{stackrel}
\usepackage{graphicx}
\usepackage{setspace}

\usepackage{threeparttable}
\renewcommand{\thetable}{\arabic{table}}
\renewcommand{\thefigure}{\arabic{figure}}

\usepackage[all]{xy}
\usepackage{tikz}
\usetikzlibrary{positioning}
\usetikzlibrary{arrows,decorations.pathmorphing,shapes}
\usepackage{tikz-cd}

\usepackage[authoryear, round]{natbib}

\usepackage[unicode=true]{hyperref}

\makeatletter

\usepackage{amsfonts}
\usepackage{pdflscape}
\usepackage{mathpazo}                 
\usepackage{microtype}                
\usepackage{bm}                       
\usepackage{listings}
\usepackage{verbatim}
\usepackage{adjustbox}
\usepackage{multirow}
\usepackage{enumitem}
\usepackage{xcolor}

\hypersetup{
    colorlinks = true,
    linkcolor  = blue,
    citecolor  = blue,
    runcolor   = blue,
    urlcolor   = blue
}

\renewcommand{\textfraction}{0.07}      
\renewcommand{\floatpagefraction}{0.85} 

\newtheorem{proposition}{Proposition}

\newenvironment{proof}{\emph{Proof.} }{$\Box$}

\AddToHook{cmd/appendix/before}{%
    \setcounter{equation}{0}%
    \setcounter{theorem}{0}%
    \setcounter{proposition}{0}%
}
\makeatother

\newcolumntype{L}[1]{>{\raggedright\arraybackslash}p{#1}}
\newcolumntype{C}[1]{>{\centering\arraybackslash}p{#1}}

\title{\vspace{-1.5cm}Getting There and Getting In: \\ How Mobility and Sorting Keep Women out of Top Startup Accelerators\thanks{The study was funded by the European Research Council (ERC Starting Grant 101221025 -- BALANCE). All errors are ours.}}
\author{Chuan Chen\thanks{Washington University in St Louis.
        Email: \href{mailto:chuan.chen@wustl.edu}{\nolinkurl{chuan.chen@wustl.edu}}.}
        \and
        Michele Fioretti\thanks{Department of Economics,
        Bocconi University, Via R\"ontgen 1, 20136 Milan, Italy, and CEPR.
        Email: \href{mailto:fioretti.m@unibocconi.it}{\nolinkurl{fioretti.m@unibocconi.it}}.}
        \and
        Junnan He\thanks{Department of Economics, Sciences Po, 27 Rue
        Saint-Guillaume, 75007 Paris, France.
        Email: \href{mailto:junnan.he@sciencespo.fr}{\nolinkurl{junnan.he@sciencespo.fr}}.}
        \and
        Yanrong Jia\thanks{Baruch College-CUNY.
        Email: \href{mailto:yanrong.jia@baruch.cuny.edu}{\nolinkurl{yanrong.jia@baruch.cuny.edu}}.}}
\date{June 19, 2026}

\begin{document}
\maketitle
\thispagestyle{empty}
\vspace*{-2.5em}


\begin{abstract}
\noindent\singlespacing
Startup accelerators are a leading gateway to venture capital, but top programs often require founders to relocate to a venture hub. From a hand-collected census of U.S.\ accelerator startups ($2008$--$2011$) followed for five years, we estimate a two-sided matching model that separates two channels behind the gender funding gap, geographic \emph{mobility} and \emph{sorting} across accelerator tiers. Women raise about $60\%$ less than men over five years; the gap concentrates among non-relocating women, is largest at active-childrearing ages, and vanishes for relocators, while the mobility cost is near zero for men. 
Removing mobility frictions raises women's match quality but not their tier; reaching the high-funding top tier also requires removing the sorting disadvantage that women face.
The $2012$ JOBS Act eased the legal barrier and capacity grew tenfold, yet the U.S.\ VC dollar gap still tripled ($2011$--$2020$): closing it needs mobility, sorting, and capacity together.
\\[1ex]
\end{abstract}

\thispagestyle{empty}
\newpage
\setcounter{page}{1}

\onehalfspacing
\section{Introduction}\label{sec:intro}

Until September~$2013$, U.S.\ securities law made physical proximity
to investors a precondition for equity fundraising from accredited
investors. Regulation~D and Rule~$506$ required private placements
to occur without general solicitation: a founder could not publicly
advertise that her company was raising capital, could pitch only to
accredited investors with whom she had a pre-existing relationship,
and had to build that relationship in person. Combined with
venture-capital concentration in California, Massachusetts, and
New York, this legal architecture made geographic mobility the
binding margin in access to outside equity capital. A startup
founded outside the three hubs that wanted a large round had to come
to the hubs.

Startup accelerators emerged in this institutional gap. Programs like
Y~Combinator (founded $2005$) and TechStars (founded $2006$) acted as
private intermediaries that bridged the geographic friction. They
brought cohorts of founders to a hub city, embedded them in an
investor network, and culminated in a curated demo day at which
graduating startups pitched to a pre-assembled audience of accredited
investors. The price of using this institutional bridge was a
relocation requirement: an admitted startup founded outside the hubs
had to move to attend.

This paper documents a large gender gap in the funding startups raise through
these programs. In a hand-collected census of every pre-policy U.S.\ for-profit
accelerator ($2008$--$2011$), women raise far less outside capital than
observably comparable men. We separate two frictions behind the gap, each
visible in a margin we observe directly. A geographic \emph{mobility} cost makes
moving to a distant hub costlier for her, so she relocates less; a
\emph{sorting} friction lowers her value to the top-tier programs, so she enters
weaker ones.

The sorting friction is where discrimination  hides. Accelerator admission is
gender-blind on its face, yet sorting is the equilibrium outcome of a selection
market, so a lower female value at the top reads equally as a higher cost of
competing there or as admission on weaker terms \citep{goldin_rouse_2000}. Both
frictions sit upstream of the investor-evaluation margin the literature
emphasizes: by the time a woman pitches to investors at demo day, mobility and sorting
have already fixed which accelerator program she pitches from.

These two frictions are complements, and that interaction is the paper's central
mechanism. A woman reaches a top-tier seat only when two conditions hold
together: she can get to the program, and she is not sorted into a lesser one.
Relieving either friction while the other binds barely closes the gap, and
easing mobility alone can even widen it, as women re-sort to better-fitting
programs that raise their probability of funding but, being lower-tier, lower the amount they raise when funded. Only the two margins together close the gap, which is
why single-margin remedies fall short: a relocation subsidy leaves sorting in
place, and a debiased admission screen leaves the top programs as far away as
ever.

The obvious remedy fares no better. The gap looks like a shortage of
top-program seats, so adding seats should erase it. The post-policy decade ran
that experiment: the $2012$ JOBS Act removed the legal mobility friction and the
top-tier accelerator industry grew roughly tenfold, yet the average gap in
U.S.\ VC deal sizes widened from \$$2$M to \$$6$M through $2020$, closing only
among deals too small to matter \citep{pitchbook,pitchbook_women}.\footnote{We
do not read the post-period causally: the legal change and the supply response
also shifted who selected into entrepreneurship and into the accelerator
pipeline, and the granular matching data the estimator needs do not exist
post-policy.} New seats, with the two frictions intact, go almost entirely to
men, so capacity is necessary but not sufficient. Closing the gap takes
mobility, tier access, and capacity together; no single lever suffices.

Three findings define the central contribution. First, conditional
on startup quality, women-founded startups raise about $60\%$ less
funding than comparable men-founded startups over five years. The
unconditional raw gap is even larger (\$$1.34$M for women versus
\$$7.03$M for men, or $81\%$ in mean five-year funding). Conditioning on
observable controls and comparing within accelerator programs absorbs
about a quarter of the unconditional gap; the residual $60\%$ shortfall
survives every empirical choice we
examined, with or without correcting for selection into programs, across
alternative functional forms, and across four inference schemes.

Second, the gap is not uniform across women. It concentrates
entirely among those who do not relocate to attend their accelerator.
Among the roughly three-quarters of female founders who stay in
their incorporation state, the funding shortfall is large and
statistically significant; among the quarter who relocate, the gap
is indistinguishable from zero. The gap is also U-shaped in founder
age: shallow in the early twenties, deepening monotonically through
the early thirties, reaching its trough around age $36$, and
partially recovering thereafter. The trough sits squarely in the
active-childrearing window for college- and graduate-educated U.S.\
women, whose median age at first birth is around $31$, and the
timing pattern mirrors the child-penalty literature on professional
women.

Third, measuring the two frictions calls for a model of the matching. Because startups and accelerators choose each other, the distance a
founder moves and the tier she lands in are outcomes of that choice, not
exogenous characteristics, so a regression on them confounds the mobility cost,
the sorting friction, and who selects into each match. A two-sided matching
model, estimated on the census, recovers both sides' preferences from the stable
assignment and separates the two frictions.

The estimated mobility cost is gender-asymmetric: the per-log-mile coefficient is
essentially zero for men but substantial for women. A relocating woman raises
about \$$3.4$ million more in cumulative five-year funding than an
otherwise-comparable non-relocator, and if all non-relocating women in our sample
could relocate, the implied welfare gain is about \$$163$ million, roughly three
times what these women collectively raised. That figure is a reduced-form upper
bound, and the counterfactual overturns its policy reading.

Removing the mobility cost alone does not close the gap, and the relocation
margin can even work against women. Freed of the per-mile penalty, a woman
re-sorts to her best-fitting program; but match value carries no top-tier
premium, so the best fit is usually a well-matched program of \emph{lower}
tier, often farther from home. Women who would otherwise move only a short way
begin moving farther, trading a higher-funded local match for a better-fitting
but lower-funded one, and the share clearing \$5M falls (by $11\%$ when we free
the distance margin alone). Easing relocation raises match quality, not funding.

Removing the sorting friction alone is no closer to sufficient. It lifts women
into the high-funding top tier, but the top programs sit far from most founders,
so a woman still has to travel to reach one. The two frictions are complements:
a woman clears a top-tier seat only if she can both reach the program and is not
steered from it. Neither lever alone moves the funding gap; together they raise
the share of women clearing \$5M by a third and put eight more women into
top-tier seats, all taken from men when capacity is fixed.

The portable contribution is conceptual. Gender gaps in entrepreneurial
financing can arise from constrained \emph{access} to high-return
intermediaries even where admission looks gender-blind on observables, because
in a selection market who reaches the top intermediaries is an equilibrium and
any discrimination is camouflaged in the sorting rather than visible in
admission rates. The gap is real and large but not a uniform glass ceiling: it
concentrates among women whose geographic mobility is constrained, and the
relocators who pay that cost show little of it.

Our setting casts access friction as two complementary halves: a
mobility cost that keeps women from reaching the top programs, and a sorting
that steers them away even when reachable, the demand-side discrimination in
latent form. Because the halves are complements, easing mobility raises a
woman's match quality but not her tier, and only removing the sorting lifts her
into the high-funding programs. The
within-tier returns penalty that investor-side interventions target is a
further margin we identify only descriptively.

These stakes reach beyond the accelerator market. Because high-return
accelerators are a pipeline into venture-backed innovation, the access gap
we document is one channel behind the broader gender gap in innovation,
where women remain under-represented among inventors largely through
differences in childhood exposure \citep{bell_etal_2019}. And because
startup formation is itself a growth margin, the gap is an efficiency
concern as much as a fairness one: closing gender gaps in access to
high-return activities raises aggregate productivity by reducing talent
misallocation \citep{hsieh_etal_2019,agte_etal_2025}.

Three institutional features make this setting unusually clean.
Admission is selective: $1$--$3\%$ acceptance rates at top programs
absorb most of the quality variation that pervades cross-VC analyses.
Programs concentrate in three coastal hubs, so $27\%$ of admitted
founders must relocate, generating the mobility variation we exploit.
Female founders enter at roughly their share in the broader VC pool
($8.6\%$ versus $11.8\%$), and admission is unrelated to any
program-level predictor we test, so the post-graduation gap arises
after admission, not through it.

The $2008$--$2011$ window predates two policy changes that altered
selection into entrepreneurship and into the accelerator pipeline:
Title II of the JOBS Act lifted the general-solicitation ban
(effective September~$2013$), and AngelList Syndicates operationalised
cross-state co-investment online. Pre- and post-policy startup
populations may therefore differ on unobservables, which is why we
restrict our analysis to the pre-policy window. The hand-collected
data are complete at the program-cohort level: we identify the $31$
for-profit U.S.\ accelerators that ran cohorts in the four-year
window, the $74$ program-cohorts they offered, and the $736$
startup-accelerator matches that resulted, observing founder
demographics, startup industry, accelerator characteristics, and
post-graduation funding at one- and five-year horizons.

\paragraph{Related literature.}
Research on the gender gap in entrepreneurial finance places the binding
margin on the \emph{investor} side, through taste, stereotypes, biased
pitch evaluation, monitoring, and unequal re-access to capital
\citep{brooks_etal_2014,kanze,guzman_kacperczyk_2019,raina2021,nguyen_etal_2024,hebert2025gender,hebert_tookes_yimfor_2025,howell_nanda_2024,calderwang_gompers_2021}.
Accelerators, the high-return intermediaries we study, bring cohorts of founders
to a hub for a few months of mentoring and network access culminating in a
presentation to investors. 

A growing literature shows accelerators shape which ventures go on to raise capital
\citep{cohenhochberg, cohen_etal_2019, gonzalezuribe_leatherbee_2018, hallen_etal_2020, yu}, though
whether they narrow or perpetuate the gender gap is itself debated
\citep{kaplan_roberts_2018, roberts2018observing, dutt_kaplan_2026, galmangodage2025not}.
We identify a margin \emph{upstream} of any investor decision: a
gender-asymmetric mobility cost that limits which high-return intermediaries
female founders can reach. The gap we focus on is one of access, not behaviour, since
women respond no less than men to incentives \citep{bandiera_etal_2021}.

This connects to a frontier literature on gender frictions in access to
high-return organisations and the misallocation of female talent, where
women are sorted into worse matches, under-promoted, and under-used
relative to their productivity
\citep[e.g.,][]{cullen_perez_truglia_2023,ashraf_bandiera_minni_2024,macchiavello_etal_2025,minni_2025,iaria_etal_2026}.
Those frictions operate \emph{within} organisations; ours sits a step
upstream, sorting female founders \emph{across} intermediaries, into
lower-tier accelerator matches, before any within-organisation decision.

The friction itself, geographic mobility tied to family, is a
well-documented source of gender gaps across labour markets, from gig work
and job search to professional careers and entrepreneurship, and underlies
both the modern glass ceiling and the ``child penalty''
\citep[e.g.,][]{bertrand2010dynamics,rosenthal2012female,goldin2014grand,azmat_ferrer_2017,bertrand_2018,kleven2019children,cook_etal_2021,le2021gender,zandberg2021family}.
Our U-shaped age gradient, troughing at active-childrearing ages, mirrors
that signature but runs through founder-program matching rather than wages
or hours.

These access gaps carry aggregate stakes: female entrepreneurship is itself
a margin of growth \citep{brush2024clearing,ashraf_delfino_glaeser_solmone_2025},
and misallocating talent away from high-return activities lowers
productivity \citep{hsieh_etal_2019,bell_etal_2019,agte_etal_2025}. We
supply a concrete micro-mechanism behind those stakes: a mobility friction
that diverts female founders from the high-return matches feeding
venture-backed growth.

Quantifying the mechanism is a matching problem, and three features of the
market select the estimator: programs admit cohorts, so matching is
many-to-one; the admission deal is standardized rather than bargained pair
by pair, so utility is non-transferable; and we observe only the matches
that formed, not the ranked
preferences submitted to a centralized clearinghouse
\citep[e.g.,][]{abdulkadiroglu_agarwal_pathak_2017}. \citet{sorensen} built an estimator
for exactly this configuration in venture capital, the closest analogue to
our setting: it recovers the match value from the pairwise-stable matches,
which is what lets us separate sorting from the downstream outcome. We tailor it to two features
of the accelerator market. First, founders cross state lines to reach a
hub, and $27.2\%$ of matches span states, so we estimate at national scale
within each six-month window rather than partitioning the market by region
as in the original. Second, we recover the match value sequentially rather
than jointly, which reuses a single match-quality residual. This sits within work on two-sided matching
estimation \citep[e.g.,][]{akkus_1stissue, agarwal_2015}, on where founders choose to locate
\citep[e.g.,][]{bryan_guzman_2026, dahl_sorenson_2012}, and on identifying quality in selection markets
\citep[e.g.,][]{chandra_etal_2016,lu2018can,cooper2022higher,fioretti_wang_2023,vatter_2025}.

The paper is structured as follows. Section~\ref{sec:legal} lays out the
institutions: the pre-policy regulatory regime and the accelerator market
it gave rise to. Section~\ref{sec:background} describes our hand-collected
data and key variables.
Section~\ref{sec:framework} develops the matching framework that
recovers unobserved match quality from observed allocations.
Section~\ref{sec:gap} documents the residual
gender funding gap and traces it to non-relocating women at
active-childrearing ages.
Section~\ref{sec:cf} quantifies the welfare cost of the friction and
identifies the capacity reallocations that close it.
Sections~\ref{sec:disc} and~\ref{sec:concl} interpret the results in
light of the accelerator industry's post-$2011$ evolution and the
$2012$ JOBS Act, and conclude.

\section{Institutions}\label{sec:legal}

Until September~2013, U.S.\ securities law made physical proximity to
investors a precondition for raising equity. Regulation~D and Rule~506(b),
the exemption most early-stage startups rely on, barred \emph{general
solicitation}. A founder could not publicly advertise a raise. She could
pitch only to accredited investors (those clearing income or net-worth
thresholds), and only to ones with whom she had a pre-existing, in-person
relationship
\citep{bradford2012, romano2012}.\footnote{The rule turned on how an
investor was reached, not on the merits of the deal. Pitching to an angel
she already knew, or being introduced to one by a mutual contact, was
permitted; the same raise posted on a public website, pitched at an event
open to the public, or cold-emailed to a stranger counted as general
solicitation and was barred. A warm introduction from someone the investor
already trusted was therefore worth more than the pitch itself.} Venture capital and accredited investors
cluster in California, Massachusetts, and New York, where roughly half of
U.S.\ accredited investors resided during our period
\citep{ivanov_bauguess_2013}. Building the relationships the law required
therefore meant being there: mobility to a hub was the binding margin in
access to outside equity.

Startup accelerators emerged in this gap as private intermediaries that
bridged the friction.

Y~Combinator (founded 2005), TechStars (2006), and their successors run
short, cohort-based programs. They bring a cohort of founders to a hub city
for about three months, take roughly a 5\% equity stake for a \$25K stipend
and intensive mentoring, and culminate in a \emph{demo day} where graduates
pitch to 50--300 accredited investors. Admission is highly selective,
1--3\% at the top programs, so the cohort is a vetted, geographically
pre-sorted entry into exactly the investor population the law required
founders to know.

The four stages of a program (announcement, application,
in-person program, demo day) are detailed in
Appendix~\ref{app:accelerator_process}. The bargain is a multi-month
relocation in exchange for that access, and that relocation requirement is
the friction this paper studies.

The market clears in semi-annual cycles: a startup applies, is admitted,
attends, and reaches demo day within roughly a six-month window, and each
program fixes its cohort once. We therefore treat each six-month window,
spanning the entire United States, as a separate matching market. Our sample
covers eight such markets between 2008 and 2011, ranging from 1 accelerator
and 17 startups in the smallest half-year to 23 and 247 in the largest, a
mean of 92 startups per market.

This architecture loosened after our period. The 2012 JOBS Act (Title~II,
effective September~2013) lifted the general-solicitation ban for verified
accredited investors, and AngelList Syndicates launched the same month to
pool investors around remote lead investors; Regulation~A+ later raised the
small-offering cap from \$5M to \$50M (2015) and Regulation Crowdfunding
opened a non-accredited channel (2016)
\citep{romano2012, agrawal2014some, knyazeva2016}. Each weakened the legal
channel through which the geographic friction operated. We study the regime
that preceded all three, when physical mobility was most binding; a
pre-/post-policy comparison would conflate the friction with two endogenous
responses, who enters entrepreneurship and who selects into accelerators
once remote capital is available, so we hold the regime fixed and exploit
cross-sectional and matching variation within the pre-policy window.

\section{Data}\label{sec:background}\label{sec:data}

We hand-collect data on every for-profit U.S.\ accelerator operating between
2008 and 2011, identified from \texttt{seed-db.com} and supplemented with
press releases, accelerator websites, and the Crunchbase, AngelList, and
CB~Insights databases.\footnote{We exclude non-profit, government, and
university-restricted programs (about 5\% of candidates) and startups whose
founder demographics or post-graduation outcomes we cannot verify (under
3\%, predominantly very early failures).} The result is 736
startup--accelerator matches across 31 accelerators running 74 cohort
programs in 20 states; cohorts average 17.7 startups (ranging from 3 to 53),
and 19 of the 74 are run by the two top-tier programs, Y~Combinator and
TechStars. Table~\ref{tab:desc} reports summary statistics for the full sample.

These are genuine venture-track startups. Graduates raise VC at 45.1\%
within a year of demo day, more than ten times the 3\% Kauffman Firm Survey
benchmark, at deal sizes (\$1.4M in 2011) on par with the broader seed/angel
market (\$0.9M); the funded sub-sample raises amounts comparable to similar
startups in the wider market (Appendix~\ref{app:vc_comparison}). The programs
cluster in the coastal hubs (36.5\% in California, Massachusetts, and New
York), and it is this concentration that activates the mobility margin:
27.2\% of admitted startups relocate cross-state to attend, a mean of about
750 miles.\footnote{We measure distance as the log of the great-circle
distance between founding- and host-state centroids, from
\citep{usCensusGeoLatLon}.}

Sixty-three of the 736 startups (8.6\%) are women-founded and 673 are
men-founded; that female share sits below the 11.8\% women-founded share
among tech startups raising their first VC over the same period.
Classifying a startup's gender is not mechanical, because founding teams
usually have several members. We determine each founder's gender from names,
photographs, and, where available, self-identification on LinkedIn or company
websites, and classify a startup as women-founded if at least one of its
founders is a woman, the standard convention in this literature
\citep[e.g.,][]{ewens_townsend_2020, gornall2024gender}.

 Other definitions are possible. A stricter convention would count only all-women teams; our
\emph{at-least-one-woman} criterion is deliberately broader, following
the literature and giving female-founded startups wider representation
in the sample.

The choice is also conservative. A mixed-gender team
can send its male founder to relocate and work the in-person investor
network, so the mobility friction binds less tightly than it does for
a team of only women, who have no such substitute. Because our sample
pools mixed and all-women teams, the gap we report is a lower bound on
the gap facing all-women startups, for whom the relocation constraint
is potentially higher.

On other observables men and women look
broadly similar (age, education, industry mix), differing sharply only on
the two dimensions we control for in the matching: engineering/science
representation (31.7\% vs.\ 67.5\%) and team size (2.02 vs.\ 2.29).

\begin{table}[t]
\centering
\caption{Descriptive statistics of accelerators, startups, and matches.}
\label{tab:desc}
\footnotesize
\begin{threeparttable}
\setlength{\tabcolsep}{6pt}
\renewcommand{\arraystretch}{0.85}
\begin{tabular}{L{6.8cm} C{2.0cm} C{2.3cm} C{2.1cm}}
\toprule
& \textbf{All} & \textbf{Women-founded} & \textbf{Men-founded}\\
\midrule
\multicolumn{4}{l}{\textbf{Panel A: Matched accelerator program (startup-level mean).}}\\
Match in startup hub (CA, MA, NY)                      & 53.5\% & 41.3\% & 54.7\% \\
Top-tier program (Y Combinator / TechStars)            & 43.3\% & 22.2\% & 45.3\% \\
Mean cohort size of matched program                    & 17.70 & 13.41 & 18.11 \\
Accelerator experience (years operating)               & 2.01 & 0.94 & 2.11 \\
Match has prior female-founder cohort                  & 42.0\% & 27.0\% & 43.4\% \\
\midrule
\multicolumn{4}{l}{\textbf{Panel B: Startup characteristics (per startup; means).}}\\
Number of startups                                     & 736 & 63 & 673 \\
Share of sample                                        & 100.0\% & 8.6\% & 91.4\% \\
Startup age at admission (years)                       & 1.76 & 1.89 & 1.75 \\
Average founder age (years)                            & 28.78 & 30.63 & 28.60 \\
At least one serial founder                            & 37.6\% & 28.6\% & 38.5\% \\
Founding-team size                                     & 2.26 & 2.02 & 2.29 \\
At least one founder with graduate degree              & 35.3\% & 42.9\% & 34.6\% \\
At least one founder with PhD                          & 7.5\% & 11.1\% & 7.1\% \\
At least one engineering/science founder               & 64.4\% & 31.7\% & 67.5\% \\
\addlinespace[0.4ex]
\textit{Industry distribution:} & & & \\
\hspace{1em}IT services                                & 38.2\% & 31.7\% & 38.8\% \\
\hspace{1em}Software                                   & 18.6\% & 22.2\% & 18.3\% \\
\hspace{1em}Data processing/hosting                    & 26.8\% & 31.7\% & 26.3\% \\
\hspace{1em}Internet \& web, other                     & 9.2\% & 3.2\% & 9.8\% \\
\midrule
\multicolumn{4}{l}{\textbf{Panel C: Match and post-graduation outcomes.}}\\
Cross-state relocation                                 & 27.2\% & 23.8\% & 27.5\% \\
\hspace{1em}Funded within 1 year                       & 45.1\% & 42.9\% & 45.3\% \\
\hspace{1em}Funded within 5 years                      & 58.3\% & 50.8\% & 59.0\% \\
\hspace{1em}$\geq\$5$M cumulative @ 5y                 & 17.4\% & 7.9\% & 18.3\% \\
\hspace{1em}$\geq\$10$M cumulative @ 5y                & 12.5\% & 3.2\% & 13.4\% \\
\hspace{1em}Mean cumulative 5y funding (\$M)           & 6.54 & 1.34 & 7.03 \\
\hspace{1em}Acquired within 5 years                    & 22.8\% & 20.6\% & 23.0\% \\
\hspace{1em}Failed within 5 years                      & 34.5\% & 44.4\% & 33.6\% \\
\bottomrule
\end{tabular}
\begin{tablenotes}\footnotesize
\item \textit{Notes.} The analytic sample contains 31 unique accelerator programs spanning 74 program-cohort instances (markets) over $2008$--$2011$. Women-founded equals at least one female founder. Funding amounts are constant 2011 dollars and exclude exit-related transactions.
\end{tablenotes}
\end{threeparttable}
\end{table}

\subsection{Gender differences}\label{sec:gender_diff}

Women and men founders differ on two distinct dimensions. The first is the
\emph{match}: which program a startup sorts into, fixed at admission and
governed by the match value, the complementarity between a startup and an
accelerator. We show the match directly affects the probability of funding over the next five years. 
The second is the \emph{funding} itself. The two are not the same outcome. 
A good match raises the \emph{odds} that a
startup clears later funding bars, but the \emph{amount} it raises is a
heavy-tailed draw only weakly tied to the match: a startup can match well and
raise nothing, or match modestly and hit a large round. Both gaps are large
here, and we trace each.

The unconditional funding gap is large and lives in the upper tail. Women
and men are about equally likely to raise any funding (one-year 42.9\%
vs.\ 45.3\%; five-year 50.8\% vs.\ 59.0\%), but women are less than half as
likely to clear \$5M (7.9\% vs.\ 18.3\%) and under a quarter as likely to
clear \$10M (3.2\% vs.\ 13.4\%); mean five-year funding is five times larger
for men (\$7.03M) than for women (\$1.34M). The match level foreshadows the
mechanism: of the 319 top-tier slots at Y~Combinator and TechStars, only 14
go to women against 305 to men, so 22.2\% of female founders match with a
top-tier program against 45.3\% of male.

That $23$-point gap in top-tier access has two sources, and they map onto two channels. The first is \emph{where} women found. Top-tier programs
sit in five states, the three coastal hubs (California, Massachusetts, New York)
plus Colorado and Washington, and being in one is close to decisive: a hub-state founder reaches
top-tier far more often than an out-of-hub one ($41.9\%$ versus $3.1\%$ among
women, $67.3\%$ versus $10.4\%$ among men). Women are less likely to found in a
hub ($49\%$ against $61\%$ of men), and that composition difference accounts for
about a third of the gap. It is the part a non-hub woman could close only by
relocating, which is what the mobility friction taxes.

The second source, the
remaining two-thirds, is the within-hub gap itself: even inside the hubs, women
reach top-tier far less than the men beside them, the $41.9\%$-against-$67.3\%$
difference above. These women are already in place and need not relocate, so
their shortfall is not about mobility; it is the sorting margin, women
under-represented in the top programs with geography held fixed, and it is what
the matching model isolates.

Because the match is two-sided, that sorting
reflects two forces: not only how women select programs, but how programs
select women. The latter is a demand-side channel, and it could reflect
discrimination.

The same gap shows up as scale. The top programs are also the large ones (24.5
startups per cohort against 9.1 elsewhere, correlated at $0.6$), so the sorting
reads on cohort size as well as tier. Women sit in smaller cohorts than men (11
versus 16), and the gap survives \emph{within} tier: among top-tier matches,
women average a cohort of 14.6 against men's 25.0. Men reach the top through
the large batches; the few women who get there are in the smaller programs.

A final cut, suggestive given the small cells, points the same way. Most
founders are of childrearing age (the median founder is 28), and among women
under 35 the ones who reach a top-tier program are disproportionately local:
20\% match cross-state, traveling 173 miles on average, against 28\% and 322
miles among childrearing-age women at non-top programs. Childrearing-age women
in top-tier programs also travel far less than the older women who reach them
(173 versus 646 miles). These numbers rest on ten childrearing-age women in
top-tier and four older, so we read the direction, not the magnitudes: when
moving is costliest, the women who reach a distant top program are the few who
did not have to move far.

\begin{table}[t]\centering\caption{Gender differences: the match and the outcome.}\label{tab:gender_diff}\small\begin{threeparttable}
\begin{tabular}{L{9cm} C{2.2cm} C{2.2cm}}\toprule
 & Women & Men\\
 & ($n=63$) & ($n=673$)\\\midrule
\multicolumn{3}{l}{\textbf{Panel A. The match (sorting).}}\\
Top-tier program (Y~Combinator / TechStars) & 22.2\% & 45.3\%\\
Founded in a hub state & 49.2\% & 61.4\%\\
Relocated cross-state to attend & 23.8\% & 27.5\%\\
Mean distance to program (miles) & 268 & 347\\
Mean cohort size & 11.0 & 16.3\\
\quad Top-tier rate $\mid$ founder in a hub state & 41.9\% & 67.3\%\\
\quad Top-tier rate $\mid$ founder outside hubs & \phantom{0}3.1\% & 10.4\%\\
\quad Mean cohort size $\mid$ top-tier match & 14.6 & 25.0\\
\midrule\multicolumn{3}{l}{\textbf{Panel B. The outcome (funding).}}\\
Raised any VC, one year & 42.9\% & 45.3\%\\
Raised any VC, five years & 50.8\% & 59.0\%\\
Cleared \$5M, five years & \phantom{0}7.9\% & 18.3\%\\
Cleared \$10M, five years & \phantom{0}3.2\% & 13.4\%\\
Mean five-year funding (\$M) & 1.34 & 7.03\\
\bottomrule\end{tabular}
\begin{tablenotes}\scriptsize\item \textit{Notes.} Means by founder gender for the 736 matched startups (63 women-founded, 673 men-founded). Panel~A describes the \emph{match}: top-tier programs are Y~Combinator and TechStars; hub states are the five hosting a top-tier program (CA, CO, MA, NY, WA); distance is the great-circle distance between founding- and host-state centroids, averaged over all founders (zero for in-state matches). The last two rows condition on whether the founder is located in a hub state. Panel~B describes the downstream \emph{funding} outcome over the five years after admission. These are population descriptives for the full matched sample, so no standard errors are reported.\end{tablenotes}\end{threeparttable}\end{table}

Table~\ref{tab:gender_diff} collects both gaps. These comparisons set up the
paper's question: how much of the gap survives once we control for
unobserved match quality?

Admission does not generate the gap. No program characteristic predicts a
startup's gender. Following the placebo logic of \citet{Giuli_Kostovetsky_2014}, a regression of the women-founded indicator on hub-state
location, top-tier status, log cohort size, and operating experience yields
no significant predictor, and adding the program's prior female-founder
share leaves it indistinguishable from zero
(Appendix~\ref{app:admission_table}). The residual gap is therefore a
within-graduate phenomenon, which is what the matching framework analyses.

\section{Empirical Framework}\label{sec:framework}

Answering these questions requires modeling the matching itself. Who matches with
which program is chosen, not assigned, so a regression of funding on program quality
and gender confounds a program's effect with selection on unobserved match quality,
and it conditions on the realized match when our questions are about how that match
would rearrange. Funding cannot resolve this from inside a single equation either:
it is realized over the years after the match forms, so it cannot enter the
matching decision, and replacing it with expected funding only reintroduces the same
selection, now to be corrected. We therefore model the matching market directly. A
first stage recovers the unobserved match quality and lets us re-solve the allocation
under each counterfactual; a second stage estimates funding net of the recovered
selection.

\subsection{First Stage: The Matching Model}

The first stage produces two objects. The structural parameters
$\boldsymbol\beta$ describe how startups and accelerators rank one another and
feed the counterfactual reallocations in Section~\ref{sec:cf}. The per-startup
residual $\hat e_s$ summarises the unobserved match quality that the observable vector 
of characteristics $\mathbf
X_{a,s}$ misses. This random variable serves as a control function
\citep{heckman1979, blundell_powell_2003} in the second-stage outcome
regressions.

\paragraph{Setup.}
The eight semi-annual markets defined in Section~\ref{sec:legal} are
indexed by $j$. Market $j$ has $M_j$ programs and $N_j$ startups; program $a$
has capacity $\bar n_a$ equal to its observed cohort size. Each market spans
the entire United States within its six-month window.

This national scope is
where our implementation differs from the original
venture-capital application, which partitions by region as well as time to keep
the simulated-likelihood computation feasible; partitioning by region would
cut the cross-state matches that we explicitly want to retain. The latent
value of pairing startup $s$ with program $a$ is
\begin{equation}\label{eq:match_value}
U_{a,s} = \mathbf X_{a,s}\boldsymbol\beta + \varepsilon_{a,s},
\end{equation}
with $\varepsilon_{a,s}\sim\mathcal N(0,1)$ an unobserved, \emph{match-specific}
component.

Equation~\eqref{eq:match_value} is the value
of a \emph{match}, not the sum of a startup's quality and a program's quality:
it measures the complementarity between a particular founder and a particular
program, how well the two go together. That is what makes the allocation worth
modelling. What the two sides match \emph{on} is the prospect of funding: a
startup and an accelerator pair up when the partnership looks likely to raise
capital, the outcome both want, so $U_{a,s}$ is in effect the pairing's
expected probability of a funded outcome, a reading Section~\ref{sec:matchgap}
confirms against the data.

If value were additive, every startup would simply rank programs by
their quality and want the same one, and there would be no sorting to explain;
the complementarity is what gives different founders different rankings. A
founder in Boston and a Boston accelerator are a strong match: nearby, same
state, easy to attend. The same founder and a California program are a weaker
match for her, and weaker still if she is raising young children, because the
cost of relocating falls on her.

The observable term $\mathbf
X_{a,s}\boldsymbol\beta$ collects the dimensions of that complementarity we can
measure, geographic proximity, same-state status, program tier and cohort size,
and how each interacts with the founder's gender; $\varepsilon_{a,s}$ is the part
the covariates miss.

The match-specific subscript is essential. A startup-level shock $\varepsilon_s$
would enter every accelerator's value for $s$ identically, so it would cancel
from every pairwise comparison and leave a startup ranking accelerators by
$\mathbf X_{a,s}\boldsymbol\beta$ alone; that structure is too rigid to
rationalise the observed sorting (the stability region it implies is empty in
most of our markets). The $\varepsilon_{a,s}$ are not $NM$ free parameters: they
are integrated out, and the matching identifies only the finite-dimensional
$\boldsymbol\beta$.

\paragraph{Identification.}\label{sec:framework_id}
OLS on equation~\eqref{eq:match_value} is infeasible because
we do not observe its left-hand side, $U_{a,s}$. The only object that
appears in our data is the matrix of realised matching $\mu$. 
We treat $\mu$ as a function that records who is paired with whom:
$\mu(s)\in\{1,\ldots,M_j\}$ is the accelerator that startup $s$ ended up at,
and $\mu(a)\subseteq\{1,\ldots,N_j\}$ is the set of startups in program $a$'s
cohort.  

The way around this identification problem
is to let $\mu$ itself identify $\boldsymbol\beta$ through pairwise
stability. For every unmatched pair $(a,s')$, at least one side must
prefer its observed partner: either $s'$ is worse for $a$ than $a$'s worst
admit, or $a$ is worse for $s'$ than $s'$'s admitted program $\mu(s')$.
Formally, one of the two inequalities
\begin{equation}\label{eq:delta_defs}
\begin{aligned}
\Delta^{(1)}_{a,s'} &\equiv \min_{s\in\mu(a)}\!\bigl(\mathbf X_{a,s}\boldsymbol\beta + \varepsilon_{a,s}\bigr) - \mathbf X_{a,s'}\boldsymbol\beta \geq \varepsilon_{a,s'}, \quad\text{or}\\
\Delta^{(2)}_{a,s'} &\equiv \mathbf X_{\mu(s'),s'}\boldsymbol\beta + \varepsilon_{\mu(s'),s'} - \mathbf X_{a,s'}\boldsymbol\beta \geq \varepsilon_{a,s'},
\end{aligned}
\end{equation}
must hold.

Each $\Delta$ is the deterministic match-value gap by which the
realized match beats the unrealized alternative, $\Delta^{(1)}_{a,s'}$ from
$a$'s side and $\Delta^{(2)}_{a,s'}$ from $s'$'s; a larger $\Delta$ means stability holds for a larger fraction of the off-equilibrium error $\varepsilon_{a,s'}$.

Since $\varepsilon_{a,s'}\sim\mathcal N(0,1)$, the pair is non-blocking with probability
$2\Phi(\max\{\Delta^{(1)}_{a,s'},\Delta^{(2)}_{a,s'}\})$, where $\Phi$ is the
standard-normal CDF. The market-$j$ likelihood for $\boldsymbol\beta$ is the product of these
non-blocking probabilities across all unmatched pairs in $j$, and the
full-sample likelihood multiplies across the $J=8$ markets.

Only pair-specific covariates contribute.\footnote{Startup-only covariates
(gender, education, age) enter $\mathbf X_{a,s}\boldsymbol\beta$ identically for
every $a$ and cancel in both $\Delta$ terms; accelerator-only covariates do the
same.} The matching therefore identifies coefficients on pair-specific variation
(distance, same-state, gender-by-mobility interactions), but not the female main
effect on funding outcomes. The latter is the second stage's job.

The same logic fixes which gender interactions we can form, and why these
four. A Female~$\times$~education or Female~$\times$~sector term is
startup-only, constant across the accelerators a given founder could
attend, and cancels exactly as the main effects do, so it is neither identified
here nor what an access friction means: education and sector describe \emph{who
the startup is}, not \emph{which programs she can reach}. The interactions that
survive are Female with the features that vary across a founder's accelerator
\emph{choices}, the two geographic-mobility margins (log-distance and
same-state) and the two accelerator-scale margins (top-tier status and
log-cohort size).\footnote{The pair-specific hub-to-hub indicator (both the
startup and the accelerator in a CA/NY/MA venture hub) enters as a main effect,
but a fifth female interaction, Female~$\times$~hub-to-hub, is not separately
identified: the simulated-likelihood estimator no longer converges to an
interior optimum and the information matrix is not positive definite. The
female coefficients are identified from the $63$ women in the sample, so a fifth
geography or tier interaction exceeds what that data allows.}

The pair-specific gender interactions therefore identify how the cost of
mobility differs by gender.\footnote{A negative coefficient on
Female~$\times$~Log-distance, for instance, would mean that an extra mile of
distance lowers the female-accelerator match value by more than the
male-accelerator one.} Mobility enters through two terms because geography has two distinct margins,
and a founder need not be deterred by both. \emph{Distance} is continuous:
Female~$\times$~Log-distance lets the per-log-mile cost of a match differ by
gender. \emph{Same-state} is discrete: it prices matching inside one's own
state, where incorporation, licensing, and local investor ties already sit, and
Female~$\times$~Same-state lets that in-state premium differ by gender.

The two are not redundant: distance is zero by construction within a state, so it
cannot price the in-state premium, and a founder can cross a nearby border at
low distance or move far within one state. Whether women are held back more by
\emph{long} moves or by \emph{leaving their state} is then an empirical
question, which the estimates of Section~\ref{sec:matchgap}
answer.\footnote{A diagnostic confirms that the estimated parameters rationalise
the observed allocation: the matches that actually occurred dominate
capacity-preserving random reshuffles, by a wider margin for women than for men,
consistent with a binding female mobility cost
(Appendix~\ref{app:vbeta_diagnostic}).}

\paragraph{Covariates.}
$\mathbf X_{a,s}$ has $p=27$ elements: twelve startup-side (female-founder
indicator, founder experience, business age, education indicators (graduate,
PhD, engineering/science), team size, founder age, and four industry dummies
for IT services, software, data processing/hosting, and internet \& web).

Four are accelerator-side (hub-state location, years of operating experience, log
cohort size, prior female-founder share).

Eleven are pair-specific (log
state-centroid distance, same-state, hub-to-hub, four founder-credential
interactions with top-tier status, and the four female interactions that
carry the paper's two margins: Female~$\times$~Log-distance and
Female~$\times$~Same-state for mobility, and Female~$\times$~Top-tier and
Female~$\times$~Log-cohort for sorting).\footnote{Top-tier status and cohort
size are related, the top programs run the larger cohorts, but they are not the
same variable: the correlation is $0.6$, and cohort size varies widely within
tier. We keep both female-by-scale interactions because each carries scale
variation the other misses.}

\paragraph{Estimation.}
We integrate $\boldsymbol\varepsilon$ out by simulation. For each market $j$
we draw $S=1{,}000$ realisations $\boldsymbol\varepsilon_j^{(t)}$ from the
prior $\mathcal N(0,I_{N_j})$, evaluate the product of non-blocking
probabilities at each draw, and average across draws:
\begin{equation}\label{eq:simloglik}
\widehat L_j(\boldsymbol\beta) \;=\; \frac{1}{S}\sum_{t=1}^{S} \prod_{(a,s')\notin\mu} 2\Phi\!\left(\max\{\Delta^{(1)}_{a,s'},\Delta^{(2)}_{a,s'}\}\,\big|\,\boldsymbol\varepsilon_j^{(t)}\right).
\end{equation}
The simulated maximum-likelihood estimator
$\hat{\boldsymbol\beta}=\arg\max_{\boldsymbol\beta}\sum_j\log\widehat L_j(\boldsymbol\beta)$
maximises the sum across markets.

Because the Monte Carlo draws inject
simulation noise, the optimum is itself a random variable that varies with
the draw set. To smooth this out we run the maximisation three times in
sequence: each run uses a fresh set of $\boldsymbol\varepsilon$ draws and
starts the optimiser at the previous run's optimum (a warm restart). The
third run's estimate is the one we report; by then it is essentially
insensitive to the particular draws.

\paragraph{Recovery of startup quality ($\hat e_s$).}
Given $\hat{\boldsymbol\beta}$, we set
\begin{equation}\label{eq:e_mean}
\hat e_s = \mathbb E\!\left[\varepsilon_{\mu(s),s} \mid \mu,\mathbf X,\hat{\boldsymbol\beta}\right],
\end{equation}
the amount of unobserved match quality the model must attribute to $s$ to
rationalise its observed match. A positive $\hat e_s$ means the model required
an above-average realized-match error $\varepsilon_{\mu(s),s}$ to make $s$'s match stable.

The posterior distribution of $\boldsymbol\varepsilon$ given the observed
matching does not admit a closed form: pairwise stability truncates the
Gaussian prior on a region whose shape depends on $\hat{\boldsymbol\beta}$
and on $\mu$ itself, so we cannot draw from it directly. We instead use
importance sampling, a standard tool when the target distribution is hard to
sample from but easy to evaluate.

The idea is to draw $T=5{,}000$
candidate vectors from the easy distribution (the prior
$\mathcal N(0,I_{N_j})$), assign each candidate a weight equal to how
well it rationalises the observed matching (the matching likelihood at that
draw), and average the realized-match error $\varepsilon_{\mu(s),s}$ across candidates using those weights.\footnote{Concretely,
$\hat e_s=\sum_t w_t\,\varepsilon_{\mu(s),s}^{(t)}\big/\sum_t w_t$
with
$w_t\;\propto\;\widehat L(\hat{\boldsymbol\beta},\boldsymbol\varepsilon^{(t)})$
and
$\boldsymbol\varepsilon^{(t)}\sim\mathcal N(0,I_{N_j})$. Draws that fall in
the stability-truncated posterior receive high weights; draws outside it
receive vanishingly small weights and effectively drop out.}

We summarise the posterior of $\varepsilon_s$ by its conditional mean
$\hat e_s$, the control-function residual, rather than by another posterior
summary such as the mode; Appendix~\ref{app:methods} discusses the choice.

\paragraph{Caveat: no portfolio complementarities.}
Identification requires that each accelerator's preference over a candidate
startup is independent of which others it admits. If programs curate cohorts
for diversity or peer effects, the matching's identification of
$\boldsymbol\beta$ is misspecified to the extent that curation correlates
with the female interactions. The admission patterns in
Section~\ref{sec:background} suggest no such curation on any observable
program-level dimension we test.

\subsection{Second Stage: Funding Outcomes}

For each post-graduation outcome $Y_s$ (e.g., a funding milestone) we estimate
\begin{equation}\label{eq:second}
Y_s
= \alpha_F\,\mathrm{Female}_s + \boldsymbol{\alpha}_Z^\prime\mathbf{Z}_s
+ \lambda\,\hat{e}_s + \mathrm{FE}_a + \mathrm{FE}_t + \eta_s,
\end{equation}
by OLS, where $\mathbf{Z}_s$ collects the startup-level controls from
$\mathbf X_{a,s}$ \emph{other than} the female indicator (founder experience,
business age, education, team size, average founder age, industry), which
enters separately with its own coefficient $\alpha_F$ so that the gender
penalty is read directly off the regression; $\mathrm{FE}_a$ and
$\mathrm{FE}_t$ are accelerator-program and year fixed effects; and
$\hat e_s$ is the matching residual from Section~\ref{sec:framework_id}. The coefficient $\alpha_F$ is the average
gender penalty conditional on selection.

\paragraph{Sequential rather than joint estimation.}
\citet{sorensen} maximises a single likelihood over the matching parameters
$\boldsymbol\beta$ and the outcome parameters
$(\alpha_F,\boldsymbol\alpha_Z,\lambda)$ at once. We instead estimate them
sequentially: recover $\hat{\boldsymbol\beta}$ and $\hat e_s$ from the
matching stage (equations~\eqref{eq:simloglik} and~\eqref{eq:e_mean}), then
plug $\hat e_s$ into the outcome regression~\eqref{eq:second}. The cost is
some efficiency relative to joint estimation; the loss is small because the
first stage is essentially insensitive to simulation noise.\footnote{The second-stage
female coefficient varies by less than $0.003$ in standard deviation across
five different $\boldsymbol\varepsilon$-draw seeds, more than an order of
magnitude smaller than the coefficient's standard error;
Appendix~\ref{app:redraws}.}

Sequential estimation also lets us reuse a
single $\hat e_s$ across the dozen outcome regressions reported below. Numerical-implementation details, including tail-stable evaluation of
the simulated likelihood, are in
Appendix~\ref{app:computation}.

The control-function role of $\hat e_s$ rests on the following result,
which generalises \citet[Proposition 2]{sorensen}.

\begin{proposition}\label{prop:selection}
Suppose the second-stage error $\eta^{as}$ and the matching-stage residual
$\varepsilon^{as}$ are jointly normal,
$$
\begin{pmatrix} \varepsilon^{as} \\ \eta^{as} \end{pmatrix}
\sim \mathcal{N}\!\left(0,\,\begin{pmatrix} 1 & \rho\sigma \\ \rho\sigma & \sigma^2\end{pmatrix}\right).
$$
Then $\mathbb{E}[\eta^{as}\mid\mu,\mathbf{X}]$ is a scalar multiple of
$\mathbb{E}[\varepsilon^{as}\mid\mu,\mathbf{X}]$, i.e., the two
expectations are collinear.
\end{proposition}

\begin{proof}
By the law of iterated expectations,
\begin{align*}
\mathbb{E}[\eta^{as}\mid\mu,\mathbf{X}]
&= \mathbb{E}\!\left[\,\mathbb{E}[\eta^{as}\mid\varepsilon,\mu,\mathbf{X}]\,\big|\,\mu,\mathbf{X}\right] \\
&= \mathbb{E}\!\left[\,\mathbb{E}[\eta^{as}\mid\varepsilon,\mathbf{X}]\,\big|\,\mu,\mathbf{X}\right] \\
&= \mathbb{E}\!\left[\,\mathbb{E}[\eta^{as}\mid\varepsilon^{as}]\,\big|\,\mu,\mathbf{X}\right] \\
&= \mathbb{E}\!\left[\,\rho\sigma\,\varepsilon^{as}\,\big|\,\mu,\mathbf{X}\right] \\
&= \rho\sigma\,\mathbb{E}[\varepsilon^{as}\mid\mu,\mathbf{X}].
\end{align*}
The second equality holds because the $\sigma$-field generated by
$(\varepsilon,\mathbf{X})$ determines the matching $\mu$. The third holds
because $(\varepsilon^{a's'},\mathbf{X})$ is independent of $\eta^{as}$
whenever $(a's')\neq(as)$. The fourth follows from the properties of the
bivariate normal distribution. Treating $\rho\sigma$ as a deterministic
parameter completes the proof.
\end{proof}

The empirical implication is direct: estimating
equation~\eqref{eq:second}, that is, regressing $Y_s$ on $\mathrm{Female}_s$,
$\mathbf Z_s$, the fixed effects, and the conditional posterior mean
$\hat e_s = \mathbb{E}[\varepsilon\mid\mu,\mathbf{X}]$, identifies $\alpha_F$
without bias from selection on unobservables, because $\hat e_s$ is a
sufficient statistic for $\mathbb{E}[\eta\mid\mu,\mathbf{X}]$. The coefficient $\lambda$ in
equation~\eqref{eq:second} estimates $\rho\sigma$, the strength of the
selection. Without $\hat e_s$, $\eta_s$ remains correlated with
$\mathbf{X}_{a,s}$ via the matching, since startups that match across long
distances tend to have unusually high $\varepsilon$ and conditioning on
their matched accelerator induces a correlation between distance and
$\eta_s$. Including $\hat e_s$ breaks this correlation and identifies
the structural $\alpha_F$.

We estimate equation~\eqref{eq:second} on three families of outcomes: (i)~probability
of clearing successive funding milestones (\$1M, \$2M, \$5M, \$10M) within one and
five years of graduation; (ii)~the log of cumulative five-year funding; and (iii)~a
five-tier ordered index combining the milestone outcomes. We estimate
(i) by linear probability and (ii) by OLS, reporting them in separate
tables; in
Section~\ref{sec:gap}~and Appendix~\ref{app:robustness} we verify that the
results are robust to alternative functional forms, including a Tobit for
the censored funding amount and quantile regressions across the
distribution.

Standard errors throughout the paper are analytic cluster-robust at
the accelerator level ($31$ clusters). Alternative constructions
(i.i.d.\ heteroskedasticity-robust, program-level clustering with
$74$ program-cohort clusters, and a restricted wild cluster
bootstrap at the accelerator level) are reported in
Table~\ref{tab:inference} and described in
Appendix~\ref{app:methods}.

\section{The Matching Estimates}\label{sec:matchgap}

Before reading the gender gap out of the matching, it is worth seeing what the
deterministic match value $\hat V_{a,s}=\mathbf X_{a,s}\hat{\boldsymbol\beta}$
captures. Startups and accelerators pair on the prospect of funding
(Section~\ref{sec:framework}), so $\hat V$ should track the \emph{probability} of
a funded outcome rather than its size. Table~\ref{tab:matchval_milestones} tests
this by sorting startups into quartiles of $\hat V$ and reading off how often
each clears the \$1M, \$5M, and \$10M five-year bars.

Across all matches the gradient is steep, and exactly where it should be. A
top-quartile startup clears the \$1M bar $43\%$ of the time against $29\%$ in the
bottom quartile (the all-matches column), but the gradient \emph{fades} as the
bar rises, to $16\%$ against $11\%$ at \$10M. Match value sorts startups into
ordinary funding success, not the rare home run.

Within the top tier the gradient vanishes (the top-tier column). At the \$1M bar
it survives, $38$ against $48\%$, but at the \$5M and \$10M bars the clearing
rate even ticks \emph{down} from the low to the high quartile, $23$ to $19$ and
$19$ to $16\%$. Those reversals are sampling noise, not a real effect: a median
split of the top tier, with more power than the quartile cells of roughly
eighty, moves \$5M-clearing by an insignificant $-2$ points (SE $5$) and
\$10M-clearing by essentially nothing. Among the elite programs, who lands a big
round is idiosyncratic.

This is what the object should do. The two sides choose each other to maximise
the chance of funding, which $\hat V$ recovers; the idiosyncratic shock that
produces a unicorn, or that separates winners inside an elite cohort, is the
residual $\varepsilon_{a,s}$ the matching integrates out and the control function
later absorbs.\footnote{Forming the full conditional-mean match value
$\hat V_{a,s}+\hat e_s$ leaves this milestone pattern unchanged (Appendix
Table~\ref{tab:matchval_milestones_eps}); $\hat e_s$ is essentially orthogonal
to $\hat V$ (correlation $-0.03$).}

\begin{table}[t]\centering\caption{Match value predicts the probability of funding, not the tail.}\label{tab:matchval_milestones}\small\begin{threeparttable}
\begin{tabular}{l c c c c}\toprule
 & & & \multicolumn{2}{c}{$\Pr(\text{clear})$, low\,$\to$\,high $\hat V$ quartile (\%)}\\\cmidrule(lr){4-5}
Milestone & $\Pr(\text{clear})$ & per-SD $\Delta$ & All matches & Top-tier only\\
 & (\%) & (pp) & & \\\midrule
\$1M   & 34 & $+6.5^{***}$ & $29 \to 43$ & $38 \to 48$\\
\$2M   & 26 & $+5.1^{***}$ & $22 \to 33$ & $31 \to 34$\\
\$5M   & 18 & $+3.6^{***}$ & $15 \to 23$ & $23 \to 19$\\
\$10M  & 13 & $+2.5^{***}$ & $11 \to 16$ & $19 \to 16$\\
\bottomrule\end{tabular}
\begin{tablenotes}\scriptsize\item \textit{Notes.} $\hat V_{a,s}=\mathbf X_{a,s}\hat{\boldsymbol\beta}$ is the estimated deterministic match value at the observed assignment. Column~2 is the share of all $736$ startups clearing each five-year funding milestone. Column~3 is the change in clearing probability per standard deviation of match value, from a linear-probability regression with standard errors clustered by program. Columns~4--5 contrast the clearing rate in the lowest and highest match-value quartile, over all matches and over top-tier matches only (the top tier has too few programs to cluster, so we report the raw spread). $^{*}p<0.1$, $^{**}p<0.05$, $^{***}p<0.01$.\end{tablenotes}\end{threeparttable}\end{table}

The matching first stage recovers how match value differs by gender, and it
reproduces from inside the model the sorting we documented in
Section~\ref{sec:gender_diff}. The four female interactions carry the gender
story; Appendix Table~\ref{tab:firststage} reports all $27$ coefficients.

\paragraph{The estimates.} All four are negative: a woman's match value falls
faster with distance, carries a smaller in-state premium, and is lower at the
top, large-cohort programs. All four are statistically significant
(Female~$\times$~log-distance $-0.58$, SE $0.008$; Female~$\times$~same-state
$-0.94$, SE $0.164$; Female~$\times$~top-tier $-0.78$, SE $0.209$;
Female~$\times$~log-cohort $-0.16$, SE $0.012$), and a joint test rejects all four
being zero.

\paragraph{Distance, not home attachment.} The same-state sign seems at first to
pull against the mobility story. A woman's in-state premium is \emph{smaller}
than a man's ($+1.4$ versus $+2.4$), so she is no more tied to her home state
than he is, if anything less; on its own, that should make women \emph{more}
willing to leave, not less. It does not, because distance swamps it almost at
once. A woman's reluctance to cross a state line overtakes a man's barely four
miles past the border, and every relocation in our data clears that threshold
many times over, the median move runs over nine hundred miles. The smaller home
premium therefore never surfaces as more moving: what keeps women close is not
attachment to home but the price of distance, which for men is essentially free.

\paragraph{Cost or discrimination?} Two mechanisms leave the same signature in
realized matches: a woman's own higher cost of relocating to or competing at the
top programs, a supply-side constraint; or those programs admitting women on
weaker terms, a demand-side preference that amounts to discrimination. The model
cannot separate them, because a pair's match value is the same object whether it
reflects a founder's cost or a program's preference.

The descriptive record
tilts toward the demand side: women resemble men on observables, no program
characteristic predicts a startup's gender at admission
(Section~\ref{sec:data}), and the largest top-tier cohorts, the scaled-up
batches, are essentially all male in this 2008--2011 window, a pattern the
leading programs themselves moved to address in later years.

We carry both
readings forward. The counterfactuals of Section~\ref{sec:cf} reallocate
startups and re-price funding from the estimated match values, which are
identical either way, so the supply-or-demand question changes the
interpretation, not the magnitudes.

\section{The Gender Funding Gap}\label{sec:gap}

This section measures the gender funding gap conditional on
accelerator-startup match quality, the residual that remains after the
matching control function absorbs selection on unobservables. Two
specifications report the same conclusion at complementary cuts of the
funding distribution: the probability of clearing successive funding
milestones, and the level of cumulative funding raised.

Table~\ref{tab:gap_prob} reports
female-founder coefficients on four binary milestones, the
\$$2$M and \$$5$M cumulative thresholds at the one- and five-year horizons.
The gap is small and statistically insignificant at the low (\$$2$M)
thresholds (columns~1 and~3: $-3.3$~pp at one year, $-6.8$~pp at five
years) and grows monotonically with the dollar cutoff: at the
\$$5$M-within-one-year threshold (column~2) it is $-2.7$~pp
(SE $1.2$), and at the \$$5$M-within-five-years threshold (column~4) it
is $-7.4$ pp (SE $4.3$).

Table~\ref{tab:gap_amount} (column~2)
extends the same pooled specification to the \$$10$M-within-five-years
threshold, where the gap reaches $-7.1$ pp (SE $2.8$). Relative to
the male baseline rates ($18.3\%$ at \$$5$M and $13.4\%$ at
\$$10$M, Table~\ref{tab:desc}), the female penalty represents a
$40\%$ shortfall at the lower threshold and a $53\%$ shortfall at the
higher one. The gap, in other words, is concentrated in the upper tail
of cumulative funding, precisely where the dollar consequences are
largest.

\begin{table}[t]\centering
\caption{The gender funding gap: probability of clearing milestones.}
\label{tab:gap_prob}\small\begin{threeparttable}
\begin{tabular}{L{5.8cm} C{2.2cm} C{2.2cm} C{2.2cm} C{2.2cm}}\toprule
& \multicolumn{2}{c}{Within 1 year} & \multicolumn{2}{c}{Within 5 years}\\
\cmidrule(lr){2-3}\cmidrule(lr){4-5}
& $\Pr(>\$2\text{M})$ & $\Pr(>\$5\text{M})$ & $\Pr(>\$2\text{M})$ & $\Pr(>\$5\text{M})$\\
& (1) & (2) & (3) & (4)\\\midrule
Female founder & $-0.033$ & $-0.027^{**}$ & $-0.068$ & $-0.074^{*}$ \\
& $(0.024)$ & $(0.012)$ & $(0.046)$ & $(0.043)$ \\
$\hat e_s$ (matching control) & $+0.018^{*}$ & $+0.012^{*}$ & $+0.015$ & $+0.011$ \\
& $(0.009)$ & $(0.007)$ & $(0.013)$ & $(0.013)$ \\
\midrule
Outcome mean (sample) & 0.064 & 0.030 & 0.253 & 0.174 \\
Founder \& industry controls & Yes & Yes & Yes & Yes\\
Year fixed effects & Yes & Yes & Yes & Yes\\
Accelerator-program fixed effects & Yes & Yes & Yes & Yes\\
Observations & 736 & 736 & 736 & 736\\
\bottomrule\end{tabular}
\begin{tablenotes}\footnotesize\item \textit{Notes.} Generated linear-probability estimates with the posterior-mean matching control, founder/industry controls, year fixed effects, and accelerator-program fixed effects. Standard errors are clustered by accelerator program. $^{*}p<0.10$, $^{**}p<0.05$, $^{***}p<0.01$.\end{tablenotes}
\end{threeparttable}\end{table}

The level results corroborate the milestone pattern.
Panel~A of Table~\ref{tab:gap_amount} reports OLS regressions of
$\log(1+\text{cumulative 5-year funding})$ and of an ordered
five-tier funding index on the female indicator and controls
(columns $3$ and $4$). Columns $1$ and $2$ extend the same pooled
specification to the \$$5$M and \$$10$M five-year milestones.

Female-founded startups raise significantly less than comparable men: the
coefficient on $\log(1+\text{5-year funding})$ is $-0.891$ (SE $0.49$,
$p<0.10$), and they place $0.29$ index points lower on the ordered five-tier
outcome (SE $0.17$). The average male-founded startup raises
\$$7.03$M over five years and the average female-founded startup
raises \$$1.34$M, an unconditional raw gap of \$$5.7$M. The
regression coefficient of $-0.891$ implies women raise
$\exp(-0.891)\approx 0.41$ of the male amount within accelerator
program and conditional on controls, a $59\%$ shortfall and an
implied conditional gap of roughly \$$4.2$M.

Panel~B re-estimates the same four specifications with a
Female~$\times$~Relocates interaction. The Female main effect
deepens to $-1.22$ log-points (SE $0.45$), and the interaction
loads at $+1.33$ (SE $0.99$): for a relocating woman the two
sum to a statistically indistinguishable $+0.11$ (SE $1.01$).
The pooled Panel~A gap is therefore concentrated among women
who did not change states, and is essentially absent among those
who did: the descriptive fact that motivates the structural
mobility analysis in Section~\ref{sec:mobility}.

\paragraph{The role of quality.}
The matching control function $\hat e_s$ enters at $+0.168$
(SE $0.104$): just shy of conventional $10\%$
significance. In standardized terms its
conditional effect on log five-year funding is about two-thirds the
female coefficient's, so it is not a vanishing covariate. The
$\hat e_s$ coefficient is essentially unchanged in Panel B once the
Female~$\times$~Relocates interaction is added ($+0.181$, SE $0.102$), and is statistically indistinguishable from zero on the
\$$5$M and \$$10$M binary outcomes in both panels.

Whether $\hat e_s$ matters for the gap is governed by its
correlations. The sample correlation between $\hat e_s$ and the female
indicator is $-0.07$, with cross-state relocation $+0.12$, and with
the Female~$\times$~Relocates interaction $-0.05$. Female and
$\hat e_s$ are therefore approximately orthogonal: adding $\hat e_s$
does not redistribute the variation associated with Female, and
re-estimating Panel B without it moves the Female~$\times$~Relocates
coefficient by under $5\%$, inside one standard error.

The shrinkage of the dollar gap, from the \$$5.7$M raw difference in means
(\$$7.03$M versus \$$1.34$M) to the \$$4.2$M implied by the $59\%$ conditional
shortfall, is therefore carried by the
within-accelerator-program comparison and the observable controls,
not by the matching control function. The matching control function
nevertheless does identifying work the within-program comparison
cannot: it captures selection on unobserved match quality, the
alternative explanation a standard OLS-on-graduates regression
cannot rule out. The female coefficient's invariance to $\hat e_s$
is therefore the identification test our two-stage design performs;
the gap survives it.

\begin{table}[t]\centering
\caption{The gender funding gap, pooled and by relocation status.}
\label{tab:gap_amount}\label{tab:mobility}\small\begin{threeparttable}
\setlength{\tabcolsep}{3pt}\begin{tabular}{L{7.2cm} C{2.2cm} C{2.2cm} C{2.2cm}}\toprule
& $\Pr(>\$5\mathrm{M})$ & $\Pr(>\$10\mathrm{M})$ & Log 5-yr funding\\
& (1) & (2) & (3)\\\midrule
\multicolumn{4}{l}{\textit{Panel A: Pooled female coefficient.}}\\
Female founder & $-0.074^{*}$ & $-0.071^{**}$ & $-0.891^{*}$ \\
& $(0.043)$ & $(0.028)$ & $(0.494)$ \\ \addlinespace[0.5ex]
\multicolumn{4}{l}{\textit{Panel B: With Female~$\times$~Relocates interaction.}}\\
Female founder (non-relocating) & $-0.095^{**}$ & $-0.083^{***}$ & $-1.221^{***}$ \\
& $(0.043)$ & $(0.025)$ & $(0.453)$ \\
Female $\times$ Relocates & $+0.082$ & $+0.048$ & $+1.333$ \\
& $(0.095)$ & $(0.084)$ & $(0.993)$ \\
\multicolumn{4}{c}{ }\\
Implied effect on relocating women & $-0.013$ & $-0.035$ & $+0.112$ \\
& $(0.095)$ & $(0.082)$ & $(1.010)$ \\
\midrule
Outcome mean (sample) & 0.174 & 0.125 & 3.94\\
Founder \& industry controls & Yes & Yes & Yes\\
Year fixed effects & Yes & Yes & Yes\\
Accelerator-program fixed effects & Yes & Yes & Yes\\
Matching control $\hat e_s$ & Yes & Yes & Yes\\
Observations & 736 & 736 & 736\\
\bottomrule\end{tabular}
\begin{tablenotes}\footnotesize\item \textit{Notes.} All three outcomes are measured over the five years after admission: columns~1--2 are the probability that cumulative five-year funding exceeds \$5M and \$10M, and column~3 is log cumulative five-year funding. Panel A pools the female coefficient (one specification per column). Panel B re-estimates the same three specifications with a Female~$\times$~Relocates interaction; the implied effect on relocating women equals the sum of the Female main effect and the interaction, with the variance computed from the full covariance matrix. Standard errors clustered by accelerator program throughout. $^{*}p<0.10$, $^{**}p<0.05$, $^{***}p<0.01$.\end{tablenotes}
\end{threeparttable}\end{table}

\paragraph{The 5-year horizon.}
The level specification raises one concern. Cumulative five-year
funding is right-skewed and bottom-censored:
$49.0\%$ of the sample receives zero VC over the horizon, so the OLS
$\log(1+\cdot)$ specification mixes the extensive and intensive
margins. Table~\ref{tab:funcform} reports five alternative
functional forms.

The Tobit coefficient, which treats zero-funding observations as
left-censored, is $-2.874$ (SE $2.098$), roughly three times the
OLS estimate of $-0.891$ in absolute value, indicating that the OLS
$\log(1+\cdot)$ specification understates the latent uncensored gap.
The full set of functional forms in
Table~\ref{tab:funcform} (Appendix~\ref{app:funcform_table})
leaves the qualitative conclusion unchanged; the distributional
structure of the gap is taken up directly in the next paragraph.

Restricting to startups that received any first-year funding
(Panel~B of Table~\ref{tab:falsification},
Appendix~\ref{app:falsification}) yields a coefficient
indistinguishable from zero on subsequent five-year amounts: women
are not less successful \emph{conditional on having been funded},
only less likely to land in the high-funding tail to begin with.

The gap measured above averages over a heterogeneous female
population. The rest of this section shows it concentrates almost
entirely among non-relocating women and peaks at the
active-childrearing ages.

\subsection{The Gap Closes for Relocating Women}\label{sec:mobility}

The gap concentrates entirely among women who do not relocate.
Panel~B of Table~\ref{tab:mobility} adds a
\emph{Female}\,$\times$\,\emph{Relocates} interaction to
equation~\eqref{eq:second}. The penalty on non-relocators is
$-1.22$ on log five-year funding (SE $0.45$, $p$-value: $<1\%$),
nearly three times Panel~A's pooled gap. The interaction is
$+1.33$ (SE $0.99$). Their sum, the implied effect on relocating
women, is $+0.11\,\log$ points, essentially zero. The same split
repeats on $\Pr(\geq\$5\text{M in }5\text{y})$: $-0.095$
(SE $0.043$, $p$-value: $<5\%$) for non-relocators against
$-0.013$ (SE $0.095$) for relocators. A specification
blind to mobility averages the gap away.

This is a sharp test of mobility-frictions. If mobility
binds, the gap disappears among women who pay the cost. If it does
not, the gap survives. Among women who clear the geographic
hurdle, the gap is indistinguishable from zero on every funding
outcome.

\begin{figure}[t]
\centering
\caption{Gender penalty and mobility return by funding quantile.}
\label{fig:quantile_decile}
\includegraphics[width=0.92\linewidth]{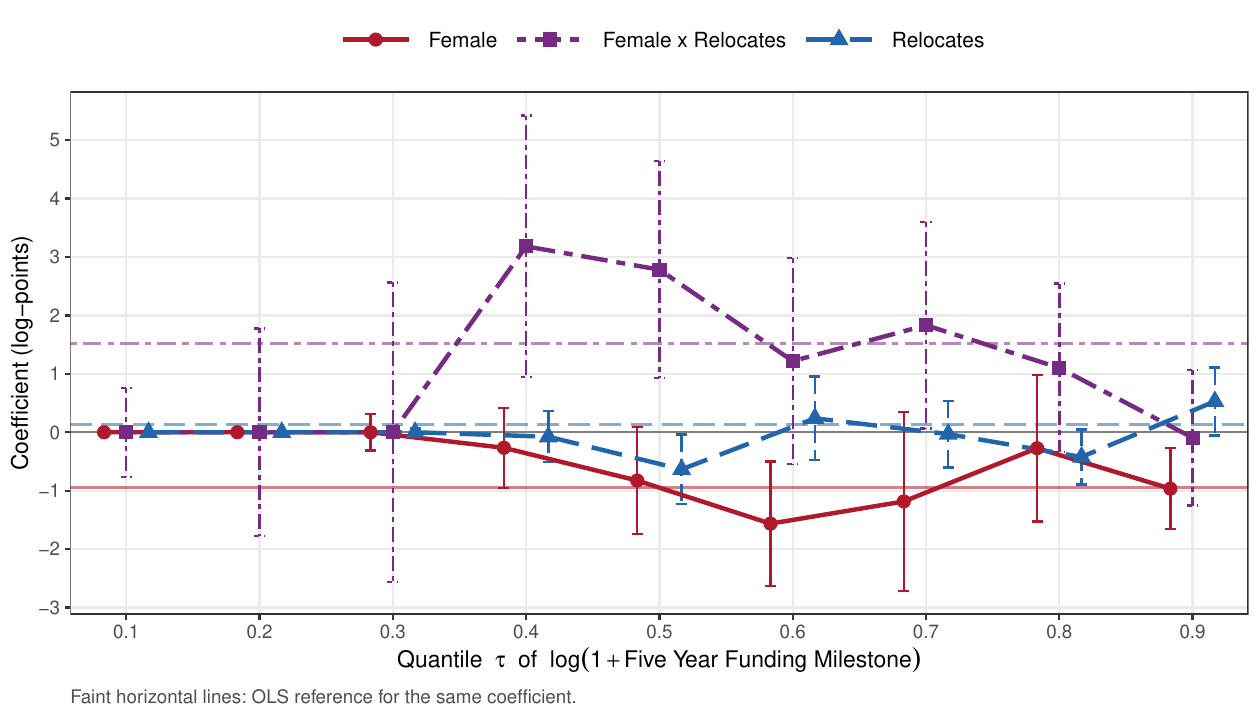}\\[0.3ex]
\parbox{0.92\textwidth}{\scriptsize \textit{Notes.} Quantile-regression coefficients at each decile $\tau\in\{0.1,0.2,\dots,0.9\}$ of $\log(1+\text{five-year funding})$, on the reduced-form quantile counterpart of Panel~B (Table~\ref{tab:gap_amount}): Female, Relocates, Female$\times$Relocates, the matching control $\hat e_s$, founder and industry controls, a top-tier accelerator indicator, log cohort size, and year dummies. Panel~B's accelerator fixed effects are dropped and replaced with the top-tier indicator and log cohort size so that the between-tier variation summed by the conditional mean stays visible in the regression. Error bars are $\pm 1$ cluster-bootstrap standard error (accelerator-level, 200 replications); bootstrap draws with $|\hat\beta|>10$ are dropped as non-converged outliers. Faint horizontal lines report the OLS coefficient on the same right-hand side.}
\end{figure}

\paragraph{Where in the funding distribution the gap lives.}
Both panels of Table~\ref{tab:mobility} include accelerator and
year fixed effects, so the $-1.22$ on non-relocators is a
within-accelerator estimate. The accelerator fixed effects absorb
the level shift between top-tier programs and the rest, where
most of the upper tail of the funding distribution lives. To see
where in the distribution the gap concentrates,
Figure~\ref{fig:quantile_decile} re-estimates the Panel~B
regression at each decile $\tau\in\{0.1,\ldots,0.9\}$. It drops
the accelerator fixed effects and replaces them with a top-tier
indicator and log cohort size, so the between-tier variation
stays in the regression and is visible across the deciles.

Three patterns emerge. Below $\tau\!\approx\!0.4$ all coefficients
are zero, because $49\%$ of the sample raises nothing. The female
penalty for non-relocators is large across the upper half: $-1.56$
at $\tau\!=\!0.6$ and $-0.97$ at $\tau\!=\!0.9$. The OLS Female
line, $-0.94$, sits next to the $\tau\!=\!0.9$ estimate. The
conditional mean is pinned by the upper tail. The Relocates main
effect is flat at zero throughout. It is identified off men, who
make up over $90\%$ of relocators and for whom moving carries no
funding consequence. Relocation bites only for women, through the
interaction.

The interaction is the central story. It peaks at the median
($+2.78$) and decays through zero by $\tau\!=\!0.9$. Through the
bulk of the distribution, relocation reveals positive selection
that closes the gap. At the top, the selection benefit gives out.
At $\tau\!=\!0.9$ the female coefficient is still $-0.97$ and the
interaction has dissipated. The implied gap for relocating women
is $-1.06$, indistinguishable from the gap for non-relocators. The
friction binds even on women whose preferences would otherwise
have carried them through.

The $+1.33$ interaction in Column (3) of Table \ref{tab:gap_amount}, the matching first-stage coefficients on
Female~$\times$~Distance, and the age gradient documented below
triangulate to a gender-asymmetric mobility friction.\footnote{One
might worry the interaction reflects self-selection on observed
outcomes. Three checks argue against this. (i) Relocation is
observed at matching, before demo-day outcomes. (ii) $\hat e_s$
is approximately uncorrelated with female age (sample correlation
$+0.03$ with age, $+0.11$ with the $28$--$38$ child-rearing band),
ruling out a quality-by-age pattern. (iii) The female-specific
first-stage coefficients on log distance ($-0.58$) and same-state
($-0.94$) are identified before any funding information enters the
model (Section~\ref{sec:framework_id}).}

\subsection{The Age Gradient}

\begin{figure}[t]
\centering
\caption{The gender funding gap is largest at active-childrearing ages.}
\label{fig:age_gradient}
\includegraphics[width=0.7\linewidth]{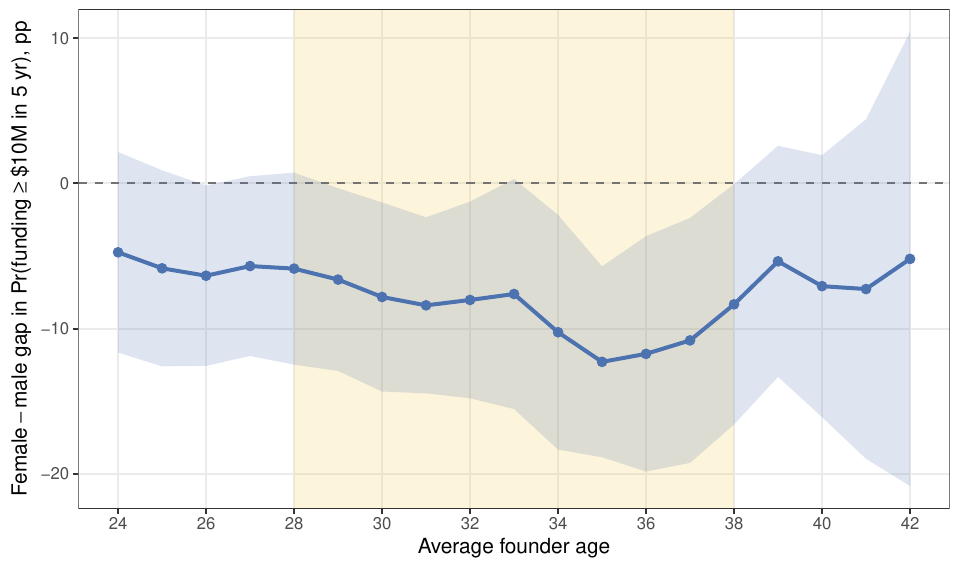}\\[0.3ex]
\parbox{0.9\textwidth}{\scriptsize \textit{Notes.} Each point is the female-founder coefficient (in percentage points) from a separate linear-probability regression of an indicator for clearing the $\geq\$10$M five-year funding milestone on a female-founder indicator, the matching control function $\hat e_s$, and founder and industry controls, with accelerator-program and year fixed effects; standard errors are clustered by accelerator program. The regression is re-estimated on a sliding $13$-year age window ($\pm 6$ years) centered at each integer average-founder age from $24$ to $42$, dropping windows with fewer than $30$ startups. The shaded band is the $90\%$ confidence interval; the vertical highlight marks ages $28$--$38$.}
\end{figure}

Figure~\ref{fig:age_gradient} plots the female coefficient on
$\Pr(\geq\$10$M raised within 5 years$)$ as a function of average founder
age, estimated separately on rolling six-year windows centered at each
integer age between $24$ and $42$. Rolling windows pool adjacent ages to
smooth simulation-noise across thin sub-cells: each window contains between
$79$ and $588$ matches, including between $10$ and $43$ female founders,
enough for informative $90\%$ confidence intervals.

The gap is shallow at the youngest ages (approximately $-5$ to $-6$
percentage points and not statistically distinguishable from zero), deepens
monotonically through the early thirties, peaks at approximately
$-12$ percentage points at age $36$, and recovers to between $-5$ and
$-8$ percentage points above age $38$. The trough sits squarely in the active-childrearing
window for college- and graduate-educated U.S.\ women, whose median age at
first birth is approximately $31$
\citep{sweeney_raley_2014,ncfmr_first_birth_2025}.\footnote{The recovery above age $38$ is imprecise
-- only $6$ female-founded startups have a team-mean age above $40$ in our
sample. We do not interpret the right-tail attenuation strongly.}

To formalise the visual trough, we estimate the female penalty
separately inside and outside the highlighted band by interacting
Female with an indicator $\mathbb{1}[28\le\text{age}\le 38]$. The
interaction coefficient is the additional female penalty at central
ages. For the $[28,38]$ band, the female penalty is $-9.6$ pp inside
($p$-value: $< 5\%$) and $-4.9$ pp outside (not significant); the
inside-outside difference of $-4.7$ pp goes in the predicted
direction but is imprecise (one-sided $p=0.15$). For a
narrower $[30,36]$ band, focused on the trough itself, the
inside-outside difference rises to $-7.0$ pp and is statistically
significant (one-sided $p=0.05$). The female penalty is therefore
concentrated at the active-childrearing ages and statistically
distinguishable from zero only there.

The trough does not reflect selection on unobserved match quality.
Within female founders, the matching control function $\hat e_s$ is
approximately uncorrelated with age (sample correlation $+0.03$
with age, $+0.11$ with the $28$--$38$ child-rearing band), so the
age signature is not an artifact of women at child-rearing ages
being matched into systematically lower-quality positions.

\paragraph{A child-penalty analogue.} The U-shaped gradient is the
matching-market analogue of the \emph{child penalty} in labor earnings
\citep{kleven2019children}, a female-specific shortfall that arrives with
the first child and eases only as the childbearing years pass.
\citet{goldin2014grand} and \citet{bertrand2010dynamics} trace that
penalty to the settings that most reward flexibility and continuous
presence; cohort accelerators are an extreme case, demanding physical
co-location for a fixed multi-month window. Here the penalty operates one
step earlier than in those studies, through \emph{which program} a founder
can attend rather than through her wages or hours: at peak-childbearing
ages her mobility cost is highest, so she takes a nearby program over a
higher-tier one, and the gap attenuates as that constraint relaxes.

\subsection{Robustness}

We close the section by asking whether the residual gap is an artifact
of how we set up the matching first stage, which startups enter the
sample, or how the standard errors are computed. None of these choices
changes the substantive conclusion.

\paragraph{Matching control function.}
The female coefficient is stable to whether $\hat e_s$ is included in
the second-stage regression, so the headline gap survives even with
the matching machinery omitted, as the low female-$\hat e_s$
correlation already implied. The matching framework is not needed to
identify the headline gap; it is needed for the mobility-mechanism
analysis in Sections~\ref{sec:mobility}--\ref{sec:cf}.

\paragraph{Subsamples.}
Table~\ref{tab:subsample} (Appendix~\ref{app:subsample_table})
re-estimates the headline specifications across four concerns:
outlier dependence on the dominant top-tier programs; geographic
concentration in versus outside startup hubs; the definition of
female founding; and the mobility prediction itself.\footnote{The
seven subsamples are: full sample (baseline); excluding
Y~Combinator and TechStars graduates; restricting to startup-hub
states (CA, MA, NY); restricting to non-hub-state matches;
all-female founding teams; non-relocating startups; and relocating
startups as a placebo.}

The gap is preserved across all restrictions that contain both genders: it
survives excluding Y~Combinator, restricting to non-hub matches, and the
non-relocating subsample (the all-female-team restriction contains no men,
so a female coefficient is not identified there). The
gap also survives the most demanding selection-on-program test
available: comparing women and men admitted to the very same
program-cohort (Appendix~\ref{app:within_program}).

\paragraph{Inference.}
Table~\ref{tab:inference} (Appendix~\ref{app:inference_table})
reassesses the female coefficient under four clustering schemes:
by startup (heteroskedasticity-robust), by program ($74$ clusters),
by accelerator ($31$ clusters, our default), plus a wild cluster
bootstrap. All four give similar standard errors. The gap is significant at $5\%$ for the
\$$10$M milestone and at $10\%$ for log funding. We adopt
accelerator-level clustering as the default.

\section{Welfare and Counterfactuals}\label{sec:cf}

The reduced-form results in Section~\ref{sec:gap}
document the symptom: women raise less than comparable men, the gap
concentrates on non-relocating women, and it peaks at active-childrearing
ages. They do not answer the policy question. Whether removing a friction
would close the gap turns on how startups would rematch to accelerators in
its absence, which women would move where and whom they would displace, and
the matching first stage delivers that reallocation. We put three questions
to it. Which of the two channels separated in Section~\ref{sec:matchgap},
the mobility friction or the top-tier sorting, drives the gap. Whether
closing it redistributes the scarce top-tier seats or needs new ones. And
how the resulting access gain compares with the within-tier penalty women
face once admitted.

We answer them by setting blocks of the four female-specific coefficients
to zero and, separately, by adding top-tier capacity. The four
(Female~$\times$~Log-distance, Female~$\times$~Same-state,
Female~$\times$~Top-tier, Female~$\times$~Log-cohort) measure how a woman's
match value responds to a candidate program's distance, same-state status,
tier, and cohort size differently from a comparable man's. Zeroing the two
\emph{mobility} coefficients, the two \emph{sorting} coefficients, or all
four is the operational meaning of removing that friction: it makes a
woman's match value respond to those features exactly as a comparable man's
does, leaving the gender-pooled component intact.

These coefficient experiments hold the top-tier seats fixed, so any seat a woman gains is one a
man loses. To ask whether the gap can instead close on \emph{new} seats rather
than redistributed ones, the second of the three questions above, and the one
the post-$2012$ capacity expansion makes concrete, we separately add $30$
top-tier slots.\footnote{About $10\%$ of the $319$ top-tier slots in our
sample, roughly one extra Y~Combinator batch ($2008$--$2011$ batches ran
$15$--$50$ startups) or two to three TechStars cohorts ($9$--$12$). Adding
more or fewer slots changes the magnitudes but none of the conclusions
below.}

\paragraph{How to read the counterfactuals.}
Panel~A of Table~\ref{tab:cf_menu} asks the same question in several ways:
holding fixed the same startups, accelerator programs, and cohort sizes, what
would the matching have looked like if one barrier had not applied to women?
The \emph{mobility} rows make women's values respond to distance and staying
in-state as men's do. The \emph{sorting} rows make women's values respond to
top-tier status and cohort size as men's do. Panel~B asks the separate capacity
question: what happens if the market has $30$ more seats, with the barriers
either left in place or removed? Across rows we then let startups and programs
rematch under the same pairwise-stability requirement used by the
estimator. The exercise is deliberately not a planner assigning founders to the
highest-funded programs; it is the original selection market rerun after one
constraint is removed or seats are added.

The simulation is anchored to the observed market. We draw the unobserved part
of match quality from its posterior conditional on the actual $2008$--$2011$
matching being stable. This conditional draw makes the simulated baseline
reproduce the data exactly: $14$ women and $305$ men in top-tier programs, and a
women-minus-men matching-utility gap of $-2.41$ per founder. Appendix~\ref{app:cf}
gives the formal stability inequalities, the matching algorithm, the
conditional draw used here, the ex-ante version that does not condition on the
observed allocation, and the planner assignment we do not use.

The three columns in the table should be read as fit, access, and money.
``Utility gap closed'' is the model's predicted match value: whether a founder
lands at a program that is a better fit for her. ``Top-tier access'' counts
women in the high-funding programs. ``Growth, clearing \$5M'' translates that
access into a funding milestone. For milestones we credit each startup the
\$1M, \$5M, or \$10M clearing rate of its own (tier,~gender) cell, so the
baseline returns the observed $18$, $6$, and $3$ women over each bar. For
dollars we credit it the cell's mean five-year funding (women raise \$$0.8$M
outside the top tier and \$$3.4$M inside, men \$$3.3$M and \$$11.5$M), so the
projected totals match observed funding exactly (\$$84.6$M for women).

This projection makes an important distinction transparent. A better
\emph{match} changes utility, but projected \emph{funding} changes only when a
woman crosses the tier boundary, because a startup that stays inside its tier
keeps the same cell rate. We report milestone growth--the rise in the share of
the $63$ women clearing a bar--and lead with it over mean dollars, which a thin
male upper tail dominates. How much a woman raises \emph{once inside} a tier,
the within-tier penalty, is a separate question we defer to
Section~\ref{sec:friction}.

\begin{table}[t]\centering\caption{Counterfactual policy menu: matching utility, top-tier access, and milestone-clearing for women.}\label{tab:cf_menu}\small\begin{threeparttable}
\begin{tabular}{l c c c}\toprule
 & Utility gap & $\Delta$ top-tier & Growth, women\\
Counterfactual & closed & (\% of female founders) & clearing \$5M\\\midrule
\multicolumn{4}{l}{\textit{Panel A. Remove the female-specific friction (fixed capacity).}}\\[0.5ex]
\quad\textit{Mobility} \;{\scriptsize(distance $+$ same-state)}      & $67\%$ & $-1.6\%$ & $-4\%$\\
\quad\ \ -- Distance \;{\scriptsize(Female~$\times$~log-distance)}   & $26\%$ & $-4.5\%$ & $-11\%$\\
\quad\ \ -- Same-state \;{\scriptsize(Female~$\times$~same-state)}   & $49\%$ & $+1.0\%$ & $+3\%$\\[0.5ex]
\quad\textit{Sorting} \;{\scriptsize(top-tier $+$ cohort)}          & $37\%$ & $+4.8\%$ & $+12\%$\\
\quad\ \ -- Top-tier \;{\scriptsize(Female~$\times$~top-tier)}       & $11\%$ & $+3.2\%$ & $+8\%$\\
\quad\ \ -- Cohort size \;{\scriptsize(Female~$\times$~log-cohort)}  & $24\%$ & $+0.7\%$ & $+2\%$\\[0.5ex]
\quad\textbf{Both channels} \;{\scriptsize(mobility $+$ sorting)}    & $97\%$ & $+12.3\%$ & $+32\%$\\
\midrule
\multicolumn{4}{l}{\textit{Panel B. Expand capacity by $30$ slots.}}\\ [1ex]
\quad -- Top-tier slots, friction kept       & $11\%$ & $-0.6\%$  & $-2\%$\\ [0.25ex]
\quad -- Non-top slots, friction kept        & $11\%$ & $-3.2\%$  & $-8\%$\\ [0.25ex]
\quad -- Top-tier slots, both removed        & $96\%$ & $+14.7\%$ & $+38\%$\\ [0.25ex]
\quad -- Non-top slots, both removed         & $97\%$ & $+12.0\%$ & $+31\%$\\ [0.25ex]
\bottomrule\end{tabular}
\begin{tablenotes}\scriptsize\item \textit{Notes.} Conditional counterfactuals on the estimated stable matching; Appendix~\ref{app:cf} defines each row and the three metrics. The simulated baseline reproduces the observed allocation: $14$ of the $63$ female founders ($22\%$) match to top-tier programs, $6$ ($9.5\%$) are projected to clear \$5M, and the women-minus-men matching-utility gap is $-2.41$ per founder. ``Utility gap closed'' is the share of that gap the reallocation removes. ``$\Delta$ top-tier'' is the change in women placed in top-tier programs, as a percentage of the $63$ female founders (baseline $22\%$). ``Growth, clearing \$5M'' is the percentage change in the number of women projected to raise at least \$5M, crediting each startup the milestone-clearing rate of its (tier,~gender) cell. The dollar projection and the within-tier penalty are detailed in Section~\ref{sec:friction}; the men's side and the ex-ante design in Appendix~\ref{app:cf}.\end{tablenotes}\end{threeparttable}\end{table}

\paragraph{Getting there, and getting in.}
Panel~A shows why the counterfactuals are not a single-margin story. Match
value and funding are different objects. Removing the mobility friction closes
$67\%$ of the match-value gap, but it barely moves the funding milestone: women
placed in top-tier programs fall by $1.6\%$ of female founders, and the number
projected to clear \$5M falls by $4\%$. The reason is mechanical but important:
mobility lets a woman re-sort to the program she values most, and that program
is often a well-matched non-top program rather than the best-funded tier.

Removing the sorting friction does the opposite. It closes less of the
match-value gap, $37\%$, but it acts directly on the tier boundary: women in
top-tier programs rise by $4.8\%$ of female founders, and the number clearing
\$5M rises by $12\%$. Mobility gets her there, to the program she would choose;
sorting gets her in, to the high-funding tier. Neither lever closes the funding
gap alone because the funding outcome requires both conditions at once.

This is the central complementarity in Table~\ref{tab:cf_menu}. For match value,
the two frictions overlap: $67\%$ plus $37\%$ exceeds the $97\%$ closure from
removing both, because either reform recovers part of the same lost fit. For
top-tier access and funding, they reinforce each other. Taken separately, the
two levers add only $+3.2\%$ of female founders to the top tier and raise \$5M
clearing by $+8\%$; taken together, they add $+12.3\%$ to top-tier access and
raise \$5M clearing by $+32\%$. A woman clears a high-funding seat only if she
can both reach it and is not steered away from it.

\paragraph{Funding lives at the high bars.}
The split arises because match value is fit, not money. A top-tier program is
better funded, but the matching model does not treat it as an intrinsically
better match for every startup. That is exactly the pattern in
Table~\ref{tab:matchval_milestones}: a one-standard-deviation better match
raises the chance of clearing \$1M by $6.5$ percentage points, but the gradient
shrinks to $2.5$ points at \$10M, and within the top tier it barely predicts
milestones at all. The matching recovers revealed fit between a startup and a
program; the high-dollar gap sits at the tier boundary and in the upper tail.

The same split appears in the counterfactuals. Zeroing the two sorting
penalties, the Sorting lever in Table~\ref{tab:cf_menu}, raises the number of
women clearing \$5M by $+12\%$. The same counterfactual, scored against the
lower \$1M bar instead, lifts clearing by only $+6\%$: \$1M turns little on
tier, so removing the tier penalty does little for it, while \$5M turns on tier
and the identical removal does twice as much. The table tracks \$5M because the
dollar gap sits there. At
fixed capacity those seats come from men; what women raise once in them, where
the gender gap survives only in the upper tail, is the within-tier penalty of
Section~\ref{sec:friction}. The men's side and the ex-ante design, which agree
in sign, are in Appendix~\ref{app:cf}.

\paragraph{Inside the mobility channel.}
The two mobility rows explain why freeing mobility need not raise funding.
Freeing the \emph{distance} penalty lets a woman re-sort to her best-fitting
program. Because fit is not the same as tier, that program is often a
well-matched non-top program; distance removal therefore raises match value but
lowers \$5M clearing by $11\%$.

Freeing the \emph{same-state} penalty works the other way. Top-tier access in
our data is overwhelmingly local: a woman who lives in a top-tier-hosting state
(California, Colorado, Massachusetts, New York, Washington) reaches a top-tier
program $42\%$ of the time, and three-quarters of those matches are in-state,
whereas a woman living elsewhere reaches one only $3\%$ of the time. Raising
women's value of an in-state match therefore pushes in-hub women into their
local top-tier program rather than a nearby lesser one, which is why the
same-state margin, alone among the mobility margins, raises rather than lowers
\$5M clearing ($+3\%$ versus $-11\%$).

The pattern mirrors the estimates of Section~\ref{sec:matchgap}: women are not
deterred from crossing a nearby state line so much as from travelling far. The
mobility lever that reaches the top tier is the one that lets an in-hub woman
stay close to a top-tier program, not the one that simply makes farther travel
less costly.

\paragraph{Inside the sorting channel.}
Sorting is also two margins, and they specialize. The \emph{top-tier} penalty
does most of the lever's funding work: zeroing it alone lifts $+3.2\%$ of women
into top-tier programs and raises \$5M clearing by $+8\%$, about two-thirds of
the combined sorting lever's $+4.8\%$ and $+12\%$. The \emph{cohort-size}
penalty alone barely moves either outcome ($+0.7\%$ and $+2\%$); it is not the
term that opens the high-funding tier.

Cohort size instead works mainly through match value. Zeroing it closes $24\%$
of the utility gap, compared with $11\%$ for the top-tier term, because the
cohort penalty lowers women's match values across many programs at once while
the top-tier penalty bites at the tier boundary. Those gains are spread through
the distribution rather than concentrated in the high-funding top tier, so they
raise fit more than dollars.

The split locates the complementarity. The two sorting margins together close
$37\%$ of the utility gap, lift $+4.8\%$ of women into the top tier, and add
$+12\%$ to \$5M clearing, each a little more than the parts would sum to
($35\%$, $+3.9\%$, $+10\%$). But the across-channel complementarity in Panel~A
runs through the top-tier margin in particular: mobility can become funding only
when the top-tier sorting term no longer steers women away from the seats they
can reach. Cohort size sits mostly on the fit margin, outside that crossing.

\paragraph{Capacity reaches women only after friction removal.}
Expanding top-tier capacity by $30$ slots while the friction remains closes
only $11\%$ of the utility gap and leaves women's top-tier count flat: in
open competition men win essentially all the new seats, about twenty of them,
because the friction holds women's match values below men's
so men out-compete them whenever seats open. This is why the untargeted
capacity the industry added after $2011$ flowed to men. Placement matters as
much as quantity: the same $30$ slots added to \emph{non-top} programs do
nothing for women either (Table~\ref{tab:cf_menu}, Panel~B), because the
binding scarcity is at the top.

Pair the capacity with friction removal and the picture inverts: women reach nine more top-tier
seats and men eleven more, both gaining, because the new capacity absorbs the
demand that friction removal unleashes instead of forcing women's gains out
of men's seats.

The two levers are complements, not substitutes: the friction
must fall for women to compete for high-return seats, and capacity must rise
for their gains not to displace men. Section~\ref{sec:disc} returns to the
policy implications.\footnote{A caveat on broader mobility policy: the $27\%$
relocation rate conceals a gender asymmetry: women relocate less but bear a
higher implicit cost when they do, so mobility-\emph{restricting} policies
(pre-NSMIA-style state securities regimes, caps on cross-state admissions) would
fall hardest on men, who relocate more, with second-order effects on women that
depend on capacity reallocation. The corresponding placebo counterfactuals
(Appendix~\ref{app:placebos}) are convention-sensitive, which is why we omit
them from the headline menu.}


\subsection{Matching Friction vs.\ Within-Tier Penalty}\label{sec:friction}

The counterfactuals above measure an \emph{access} barrier: which tier a woman
reaches. The remaining funding gap can also reflect a \emph{returns} barrier:
how much she raises conditional on being in a given tier. This subsection puts
the two on the same scale. The matching friction moves women across tiers; the
within-tier penalty prices what women would raise if, within each tier, they
received men's funding outcomes.

The access effect is the counterfactual in Table~\ref{tab:cf_menu}. Removing
both female-specific matching frictions at fixed capacity moves about eight
women into top-tier programs and is worth about $\$22$M to women under the
cell-mean projection. That figure comes only from tier crossings: a woman who
stays outside the top tier keeps the non-top female mean, and a woman who
crosses into the top tier receives the top-tier female mean.

The within-tier benchmark asks a different question. Holding each woman's tier
fixed, assign her the male mean funding in that tier. At the mean this adds
$\$239$M, roughly ten times the access effect. But this is an upper-tail number,
not a typical-founder number. Funding is extremely right-skewed, and the male
mean is pulled up by a small number of very large rounds.

Several checks show how concentrated that benchmark is. Trimming the top
$5\%$ of outcomes cuts the within-tier figure to $\$110$M. Conditioning on
observable startup and program characteristics lowers it further, to about
$\$66$M, roughly $\$1$M per woman, and the associated within-tier female
coefficient is statistically insignificant ($-0.58$ log points, SE $0.58$;
Appendix~\ref{app:within_tier} reports the full set of bounds).

The median tells the same story more starkly. The typical founder bears no
within-tier penalty: the median top-tier woman raises $\$1.4$M, above the
median top-tier man at $\$0.5$M, and the non-top median is zero for both
genders. Milestone rates inside the top tier also match or favor women at the
lower bars: women clear \$1M and \$5M at rates of $57\%$ and $29\%$, compared
with men's $45\%$ and $25\%$. The within-tier gap appears only at the largest
rounds, where women trail at the \$10M bar ($14\%$ versus $20\%$).

Thus the two forces are complements, not substitutes. The matching friction is
an access barrier: it explains why women are under-represented in high-return
programs. The within-tier penalty is a returns barrier: it explains why, among
the rare very large outcomes, women remain scarce. For the typical founder the
returns barrier is small or absent; in the upper tail it can dominate the mean.
Closing the funding gap therefore requires both access to the high-return tier
and returns within that tier, while the matching counterfactual reaches only the
first.

\section{Discussion}\label{sec:disc}

That the gap sits at \emph{access}, which tier a woman reaches, does not make
it free of discrimination. Admission looks gender-blind on observables
(Section~\ref{sec:data}), but in a selection market discrimination need not show
up there: it is absorbed into the sorting, surfacing as women's lower match
value at the top-tier programs. Whether that reflects a higher cost of competing
at the top or admission on weaker terms the market hides, and our counterfactual
removes it either way.

The access barrier has two halves, and they are complements. Easing women's
\emph{mobility}, the cost of relocating to a distant top program, lets them
sort to far better-matched programs and closes almost the entire
matching-utility gap, but it raises match \emph{quality} without raising
\emph{tier}: a freed woman re-sorts to a better-fitting program, usually a
well-matched non-top one nearer home.

Lifting her into the high-funding top tier
takes removing the second half, the \emph{sorting} that steers women from the
leading programs, the camouflaged discrimination above. Neither half alone moves
the funding gap; together they move about eight women into top-tier seats and
raise the number clearing \$5M by a third (Section~\ref{sec:cf}). A gap that
looks like a relocation gap thus closes only when the relocation cost \emph{and}
the top-tier sorting fall together, the central complementarity of the paper.

Capacity's role is narrower than it first appears. It does not open the tier to
women (the frictions do that); it determines only \emph{who pays} for the seats
they win. At fixed capacity the eight seats women gain come from men
one-for-one, a redistribution worth \$$22$M to women against \$$62$M to men;
expanding capacity lets both sides gain, but capacity added while the friction
remains flows almost entirely to men, who out-compete women for the new seats
(Table~\ref{tab:cf_menu}, Panel~B).

Untargeted expansion, the closest analog to
what the industry delivered, is the lever that fails. The larger dollar sums sit
not in this access reallocation but in the within-tier returns gap, a
tail-driven \$$239$M at the mean that is near zero for the typical founder
(Appendix~\ref{app:within_tier}). It is a lesson of the second best: relieving
one distortion need not help while another binds
\citep{lipsey_lancaster_1956,hsieh_moretti_2019}.

The decade since our sample is, in effect, a test, and it bears the diagnosis
out. The $2012$ JOBS Act lifted the general-solicitation ban\footnote{Legal and regulatory design shapes
entrepreneurial entry and access on margins beyond this one: \citet{guzman_2025}
finds that modernizing state corporate law raised business formation, with the
largest gains for women and Black founders, while \citet{ewens_farremensa_2020}
show that loosening private-placement rules let firms raise outside capital
privately and stay private longer.} and AngelList
Syndicates operationalised remote co-investment, dissolving the legal friction,
the same-state intercept (Section~\ref{sec:matchgap}), but not the binding
per-mile distance cost, which, like the top-tier sorting, was left in place. In parallel the top-tier industry grew an order of
magnitude, Y~Combinator from roughly $30$ startups a batch to $200$--$400$ and
TechStars from one Boulder programme to a global network, adding precisely the
capacity that, without friction removal, flows to men.

The diversity initiatives of the period acted on neither binding margin: the Female Founders Fund ($2014$)
and Backstage Capital ($2015$) \emph{invested} in women-led startups, Project
Diane ($2016$) \emph{measured} their under-representation, and All~Raise
($2017$) and Y~Combinator's Female Founders Conference \emph{convened} and
mentored them, supplying capital and visibility to women already in the market,
but lowering no founder's cost of moving to a hub and changing no program's
admissions.

\begin{figure}[t]
\centering
\caption{The gender gap in U.S.\ venture-capital funding, 2011--2020.}
\label{fig:gap}
\captionsetup[subfigure]{skip=2.5ex,font=footnotesize}
\begin{subfigure}[b]{0.46\textwidth}
\centering
\includegraphics[width=\linewidth]{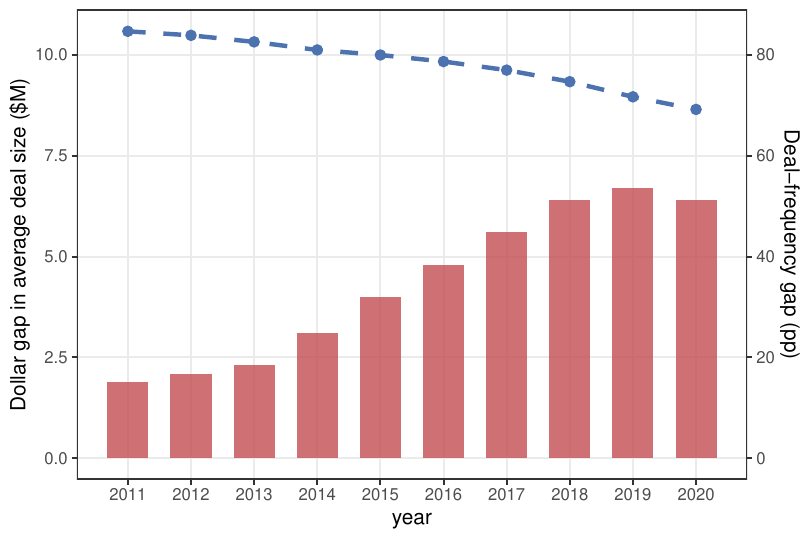}\par\medskip
\caption{Dollar gap widens, frequency narrows.}
\label{fig:gap_a}
\end{subfigure}\hfill
\begin{subfigure}[b]{0.46\textwidth}
\centering
\includegraphics[width=\linewidth]{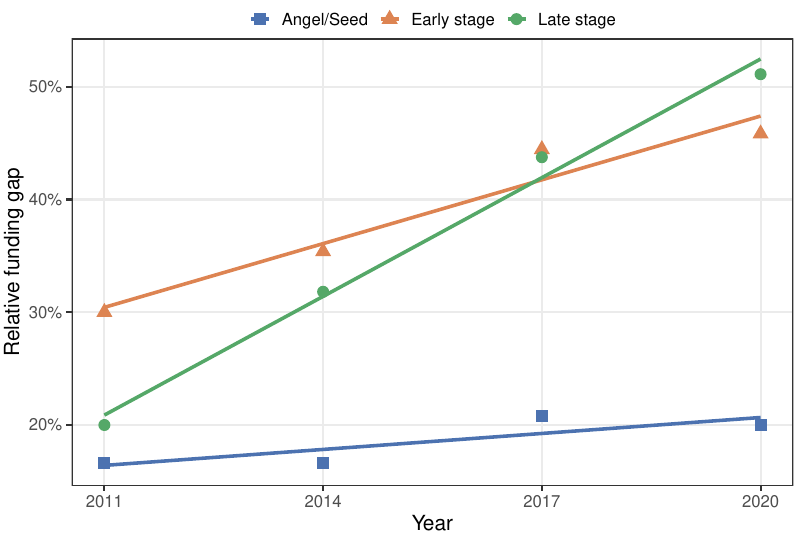}\par\medskip
\caption{Gap rises in early- and late-stage deals.}
\label{fig:gap_b}
\end{subfigure}
\\[1.2ex]
\parbox{\textwidth}{\footnotesize \textit{Notes.} Author-tabulated PitchBook annual aggregates; source data \citep{pitchbook,pitchbook_women}. Panel (a) plots the dollar gap in mean deal size (red bars, right axis) and the deal-frequency gap (blue dashed line, left axis). Panel (b) plots $1 - (\text{women-founded mean})/(\text{all-male-founded mean})$ by VC stage with linear-trend fits.}
\end{figure}

The aggregate gender gap in VC deal sizes nonetheless \emph{widened}
from \$2M to \$6M between $2011$ and $2020$ (Figure~\ref{fig:gap}); since startup
formation is a growth margin, misallocating female talent on this scale is also
a first-order loss of aggregate productivity
\citep{hsieh_etal_2019,agte_etal_2025}.

Our levers are different. The mobility lever is geographic: on-site childcare,
family-housing stipends, hybrid formats, and the multi-city satellite cohorts
TechStars pioneered cut the cost of relocating for the founders with young
children for whom it binds hardest. The sorting lever is procedural: structured,
less discretionary admission and evaluation, which the acceleration-and-bias
literature finds can shrink demand-side steering without a quota
\citep{kaplan_roberts_2018}.

The one shock that has reached the first of these
came late: the move to fully remote cohorts in $2020$ let a founder join a top
batch without moving her family, and the first all-remote batches report a
higher women-founder share.\footnote{Press coverage of Y~Combinator's
Summer~$2020$ batch, its first fully online, put women-founded companies at
about $16\%$ against $10$--$12\%$ earlier in the decade; we read this as
suggestive pending firm-level data.} Coinciding with the pandemic and the
industry's own diversity push, we read it as motivating the mechanism, not
settling it.

Modelling that post-policy environment, with program format and entrepreneurial
selection made endogenous and the accelerator treated as a mission-oriented
intermediary \citep{besley_ghatak_2005,hart_zingales_2017,fioretti2022caring},
is the natural next step. The contribution here is to identify the mechanism in
the cleanest institutional window the U.S.\ accelerator industry has offered,
and to show that the levers it implicates, relief of the founder-mobility
margin and of the top-tier sorting, are distinct from the legal and
supply-side levers already pulled.

\section{Conclusion}\label{sec:concl}

Women who graduate from U.S.\ startup accelerators raise about $60\%$ less
venture capital than comparable men over five years, even though admission is
approximately gender-blind on observables. The gap is not uniform. It is
concentrated among women who do not relocate to attend their accelerator, is
deepest at active-childrearing ages, and disappears among women who do relocate.
A two-sided matching model, estimated on the pre-policy accelerator market,
traces this pattern to a gender-asymmetric mobility cost. But the
counterfactuals show why mobility alone is not enough. Removing the mobility
friction lets a woman reach the program she values most; it does not necessarily
move her into the high-funding top tier. Funding requires a second margin:
removing the sorting that steers women away from leading programs. And because
top-tier seats are scarce, access gains at fixed capacity come from men unless
capacity expands as well. Mobility gets her there, sorting gets her in, and
capacity determines whether the gain is zero-sum.

The broader lesson is that gender gaps in entrepreneurial finance can arise
upstream of investor evaluation, through unequal access to high-return
intermediaries. In such markets, discrimination need not appear as a gender gap
in admission rates. It can be embedded in equilibrium sorting: who can reach the
intermediary, who is valued there, and who is diverted to a lower-return path.
This distinction matters for policy. The $2012$ JOBS Act and the end of the
general-solicitation ban lowered the legal cost of reaching investors, while the
accelerator industry expanded sharply afterward. The fact that the funding gap
widened over the same period is consistent with our model's warning that legal
portability and capacity growth need not close the gap if founder-level mobility
costs and top-tier sorting remain binding. Closing gaps in entrepreneurial
finance therefore requires attention not only to how capital is allocated after
founders enter the market, but also to the institutions that determine which
founders reach the highest-return platforms in the first place.

\begingroup
\setlength{\bibsep}{7pt} 
\bibliographystyle{abbrvnat}
\bibliography{references}
\endgroup

\clearpage
\appendix
\renewcommand{\thepage}{OA-\arabic{page}}
\setcounter{page}{1}
\section*{\Huge Online Appendix}
\addcontentsline{toc}{section}{Online Appendix}
\renewcommand{\thetable}{\thesection\arabic{table}}
\renewcommand{\thefigure}{\thesection\arabic{figure}}
\makeatletter
\@addtoreset{table}{section}
\@addtoreset{figure}{section}
\makeatother

\setlength{\textfloatsep}{26pt plus 6pt minus 2pt}
\setlength{\floatsep}{24pt plus 6pt minus 2pt}
\setlength{\intextsep}{24pt plus 6pt minus 2pt}
\renewcommand{\floatpagefraction}{0.6}
\renewcommand{\textfraction}{0.15}
\section{Accelerator Process and Admission Timeline}\label{app:accelerator_process}

The typical for-profit accelerator program in our sample period proceeds
through four sequential stages. The first is the public announcement.
Each accelerator publishes program details on its website and through
press releases: cohort size, location, equity stake, program dates, and
application deadline. These terms rarely change post-announcement and
are not subject to negotiation with applicants. The second stage is
application and selection. Startups submit applications to the
accelerators they would like to join. Programs in our sample receive
between roughly $200$ and $3{,}000$ applications per cohort, with
Y~Combinator at the upper end. Accelerator partners and selected mentors
review applications and conduct in-person interviews, and admissions are
decided based on predetermined cohort capacities set at program launch.

The third stage is the program itself. Admitted startups attend in
person at the accelerator's location for a fixed period, typically three
months. The accelerator provides mentorship, network access,
business-model coaching, and in most cases a modest seed-capital stipend
of approximately \$$25{,}000$ in exchange for the equity stake. The
fourth and final stage is demo day. The accelerator hosts an event at
which graduating startups pitch to a curated network of $50$ to $300$
venture-capital and angel investors. Post-demo-day, startups are free to
raise capital from any source; in practice, investors who attended the
demo day are disproportionately represented in graduates' immediate
fundraising.

This four-step timeline makes the accelerator a clean institutional setting
for our analysis. Admission decisions are made at the program-launch stage
based on observable startup characteristics (and unobservable characteristics
that are correlated with VC preferences, which we model via the matching
control function). Post-graduation funding outcomes reflect both the
investor-side selection and the founder-side mobility friction that we
identify.

\paragraph{Admission patterns.}\label{app:admission_table}
Table~\ref{tab:admission} reports OLS regressions of the
women-founded indicator on accelerator-program characteristics
(hub-state location, top-tier status, log cohort size, and years of
operating experience); column 2 adds match-year fixed effects and
column 3 adds the program's prior female-founder share. None of the four program-level
predictors moves the woman-founded indicator by more than two-tenths
of a standard deviation, and the prior-share coefficient itself is
$-0.008$ (SE $0.081$), indistinguishable from zero. Discussion in
Section~\ref{sec:background}.

\begin{table}[!htbp]\centering\caption{Accelerator admission patterns by gender.}
\label{tab:admission}\small\begin{threeparttable}\begin{tabular}{L{5.4cm} C{1.7cm} C{1.7cm} C{1.7cm}}\toprule
& (1) & (2) & (3)\\\midrule
Accelerator in startup hub (CA / MA / NY) & $-0.019$ & $-0.022$ & $-0.020$ \\
& $(0.029)$ & $(0.029)$ & $(0.028)$ \\
Top-tier program (Y~Combinator / TechStars) & $-0.025$ & $-0.014$ & $-0.015$ \\
& $(0.034)$ & $(0.036)$ & $(0.034)$ \\
Log cohort size & $+0.030$ & $+0.031$ & $+0.032$ \\
& $(0.046)$ & $(0.049)$ & $(0.046)$ \\
Accelerator experience (years) & $-0.022^{*}$ & $-0.024^{*}$ & $-0.024^{*}$ \\
& $(0.013)$ & $(0.014)$ & $(0.014)$ \\
Program female-founder share & n.a. & n.a. & $-0.008$ \\
&  &  & $(0.081)$ \\
\midrule
$N$ & 736 & 736 & 736 \\
Mean of dependent variable & $0.086$ & $0.086$ & $0.086$ \\
\bottomrule\end{tabular}\begin{tablenotes}\scriptsize\item \textit{Notes.} Linear-probability regressions of an indicator for ``startup is women-founded'' on accelerator-program characteristics, estimated on the analytic sample of $N=736$ startup-accelerator matches over $2008$--$2011$. Column (1) includes hub-state location, top-tier-program status, log cohort size, and years of operating experience; column (2) adds match-year fixed effects; column (3) adds the accelerator's prior female-founder share. Standard errors clustered at the accelerator-program level. $^{*}p<0.10$, $^{**}p<0.05$, $^{***}p<0.01$.\end{tablenotes}\end{threeparttable}\end{table}

\section{Comparison with the Broader VC Market}\label{app:vc_comparison}

This appendix compares VC outcomes for accelerator graduates in our sample with
those of comparable startups in the broader U.S.\ venture-capital market. The
purpose is to support two assumptions used in the main text: (i) that
accelerators are a representative quality filter for early-stage startups,
not a unique pool; and (ii) that the gender measurement we use in our
accelerator sample maps onto the gender measurement used in the broader VC
market.

\paragraph{Funding levels.}
In 2008, accelerator graduates raised on average
\$0.82M of VC within one year of demo day; comparable seed/angel-stage VC
deals across the same year averaged \$1.06M. Comparable averages for 2009,
2010, and 2011 are \$1.37M vs.\ \$1.00M, \$1.36M vs.\ \$0.86M, and \$1.45M
vs.\ \$0.87M, respectively. Accelerator graduates received slightly larger
deals than the seed/angel benchmark in 2009--2011 and indistinguishable
amounts in 2008. Five-year cumulative funding (averaged per-funded-startup,
divided by 5) for accelerator graduates ranged from \$1.55M to \$6.70M
(annualized), comparable to a synthetic seed-cohort benchmark of
\$1.95M--\$2.64M (annualized). Source: Authors' calculations from
\citep{pitchbook} and the Kauffman Firm Survey.

\paragraph{Industries.}
All accelerator graduates in our sample are in tech;
the average tech-industry VC deal size in 2008--2011 was \$5.49M--\$7.63M.
Within software, average market deal sizes were \$4.41M--\$5.79M. Accelerator
graduates that secured five-year cumulative funding of \$5M+ are therefore
performing at industry-average levels.

\paragraph{Gender shares.}
As reported in Section~\ref{sec:background}, female
founders comprise $8.6\%$ of accelerator participants in our sample versus
$11.8\%$ of women-founded tech startups receiving first VC deals over the
same period \citep{pitchbook_women}. Accelerators were not preferentially
admitting women in our sample period.

\paragraph{Implications.}
The accelerator sample is best understood as a
quality-screened subset of the broader early-stage VC pipeline, with funding
trajectories that approximate but do not exceed the seed/angel-stage VC
average. This supports the framing of accelerators as a useful laboratory for
within-industry comparisons rather than as a unique sub-market.

\section{Robustness of the First Stage}\label{app:robustness_first}

This section reports robustness checks on the matching first stage: the full coefficient estimates, their stability across simulation redraws, and a diagnostic that the estimated equilibrium rationalises the observed allocation.

\paragraph{First-stage matching coefficients.}\label{app:firststage_coefs}
Table~\ref{tab:firststage} reports the full set of first-stage
matching coefficients estimated by maximum simulated likelihood on
the $27$ covariates of the headline specification, with analytical
Hessian standard errors. The four female-specific
pair coefficients are the matching parameters that the
\textit{Frictionless world} and \textit{Frictionless world at fixed
capacity} counterfactuals of Section~\ref{sec:cf} set to zero.

\begin{table}[!htbp]\centering\caption{First-Stage Matching Estimates and Counterfactual Utility Change.}
\label{tab:firststage}\footnotesize\begin{threeparttable}\setlength{\tabcolsep}{4pt}
\begin{tabular}{L{6.2cm} C{1.8cm} C{1.8cm} C{1.8cm}}\toprule
\multicolumn{4}{l}{\textbf{Panel A: First-stage matching coefficients.}}\\
\addlinespace[0.3ex]
& $\hat\beta$ & Hessian SE\\\midrule
Log distance (state centroids, miles + 1) & $-0.011$ & $(0.011)$\\
Same state & $+2.357$ & $(0.053)$\\
Both startup and program in CA/MA/NY hub & $+0.135$ & $(0.074)$\\
Female $\times$ Log distance & $-0.576$ & $(0.008)$\\
Female $\times$ Same state & $-0.944$ & $(0.164)$\\
Female $\times$ Top-tier program & $-0.779$ & $(0.209)$\\
Female $\times$ Log cohort & $-0.164$ & $(0.012)$\\
Engineering/Science founder $\times$ Top-tier & $+0.451$ & $(0.076)$\\
Software industry $\times$ Top-tier & $-0.161$ & $(0.104)$\\
PhD founder $\times$ Top-tier & $+0.379$ & $(0.142)$\\
No first-time founder $\times$ Top-tier & $-0.368$ & $(0.075)$\\
Female founder (level) & $+0.012$ & $(0.065)$\\
Average founder age & $-1.895$ & $(0.451)$\\
Log cohort size & $+0.426$ & $(0.055)$\\
Accelerator experience years & $+0.182$ & $(0.016)$\\
Accelerator in CA/MA/NY hub & $-0.047$ & $(0.065)$\\
\midrule
Number of total covariates & 27\\
Number of observed matches & 736\\
Estimation & MSL\\
\midrule
\multicolumn{4}{l}{\textbf{Panel B: Matching utility, per founder.}}\\
\addlinespace[0.3ex]
& Women & Men & Gender gap \\\midrule
Baseline (mean $V$ at observed assignments) & $1.01$ & $3.42$ & $-2.41$ \\
\addlinespace[0.4ex]
\multicolumn{4}{l}{\textit{Change in mean $V$ under each counterfactual:}}\\
\quad \textit{Frictionless world} (A) & $+2.92$ & $+0.60$ & $+2.33$ \\
\quad \textit{Frictionless world at fixed capacity} (B) & $+0.82$ & $+0.22$ & $+0.60$ \\
\quad \textit{Targeted capacity expansion} (C) & $-0.27$ & $+0.00$ & $-0.27$ \\
\bottomrule\end{tabular}
\begin{tablenotes}\scriptsize\item \textit{Notes.} Standard errors (in parentheses) are analytical Hessian standard errors from the numerical Hessian of the simulated log-likelihood at the optimum. Panel~B reports the mean per-founder change in matching utility $\hat V_{a,s} = \mathbf{X}_{a,s}^\prime\hat{\boldsymbol\beta}$ across the three counterfactuals of Section~\ref{sec:cf}, decomposed by gender. Frictionless world (A): each startup picks the program that maximizes its match value under friction-free coefficients. Frictionless world at fixed capacity (B): the matching linear program with the four female-by-pair coefficients zeroed and cohort sizes held at observed values. Targeted capacity expansion (C): $30$ new top-tier slots reserved for non-top-tier women, valued at the within-tier women's mean utility (the female-specific friction is not removed). The metric is projection-free: it does not pass through a dollar-funding regression. The matching utility itself is identified up to a positive affine transformation, so the signs of $\Delta V$ and the ratios across rows (between counterfactuals and between genders) are normalization-invariant, while absolute magnitudes are not.\end{tablenotes}
\end{threeparttable}\end{table}

\paragraph{Across-redraw stability of the estimates.}\label{app:redraws}
The matching first stage runs three outer iterations of the
maximum-simulated-likelihood optimisation, each with fresh
$\boldsymbol\varepsilon$ draws and a warm restart from the previous
iteration's optimum. Across these redraws the female-specific
pair-interaction coefficients keep stable signs in every iteration,
though magnitudes vary by about $\pm 30\%$ around their final values;
the non-pair-specific same-side coefficients are noisier (average
founder age in particular flips sign across redraws), reflecting that
those parameters are identified mostly through interactions with the
accelerator side and so absorb simulation noise once explicit
interactions enter the specification. We report point estimates at the
locked-in iteration with NSim$=5{,}000$.

The second stage is far more stable. Re-estimating the full pipeline
(matching first stage with fresh $\boldsymbol\varepsilon$ draws and warm
restarts; second-stage outcome regression) under five different random
seeds, the female coefficient on $\log(1+\text{5y funding})$ ranges from
$-0.925$ to $-0.932$ (SD $0.003$) and the mobility-interaction
coefficient from $+1.195$ to $+1.281$ (SD $0.037$). This across-redraw
simulation noise is an order of magnitude smaller than the
coefficient's standard error, validating our reporting of
point estimates from a locked-in simulation seed.

\paragraph{A reshuffling check on the estimated match values.}\label{app:vbeta_diagnostic}
The matching estimator delivers a deterministic match value
$V_{a,s}=\mathbf{X}_{a,s}\hat{\boldsymbol\beta}$ that pairwise
stability requires the observed allocation to maximise. We verify
this by re-shuffling each startup to an alternative accelerator
within its market ($200$ uniform random draws per startup, sampled
with replacement and without re-imposing the original capacity
constraints) and computing per-startup $V_{a,s}$ under both
matches with the estimated matching coefficients. Across the $736$
startups in the diagnostic, mean per-startup $V_{a,s}$ is $+3.21$
under observed matches and $+1.30$ under random matches, an
average observed-minus-random surplus of $+1.92$. The observed
allocation beats the random one for $85.5\%$ of startups.

The wedge binds particularly hard along the dimensions captured by
the gender-asymmetric pair coefficients
($\hat\beta_{\text{Female}\times\text{Log-distance}}=-0.58$,
$\hat\beta_{\text{Female}\times\text{Same-state}}=-0.94$, plus the
additional Female~$\times$~Top-tier penalty), which enter
$V_{a,s}$ directly: random reshuffles disrupt the very mobility
and same-state dimensions on which the equilibrium has been
forced to economise for women.

The matching's $V$ already internalises the gender-asymmetric mobility
cost as a disutility, so women's equilibrium $V$ is achieved at
matches that minimise their distance and same-state penalties. The
funding regression, by contrast, identifies the \emph{causal} effects
of those characteristics on cumulative funding under the matching
control function of Proposition~\ref{prop:selection}, and those
effects do not embed the mobility cost as a disutility. The wedge
between the constrained-equilibrium $V$ and the unconstrained
allocation that the causal-coefficient regression implies is the
welfare cost of the friction; the diagnostic confirms that the wedge
arises from a binding constraint, not from a mis-specified equilibrium.

We do not, however, use $V$ directly to compute welfare in
Section~\ref{sec:cf} because in our data the matching's deterministic
value is weakly predictive of funding outcomes (a regression of log
funding on $\hat V$ alone returns an insignificant
coefficient).

\paragraph{Match value and funding milestones.} While the deterministic match
value is only weakly related to the dollar \emph{amount} raised (above), it
tracks the \emph{probability} of clearing successive funding milestones, as
Table~\ref{tab:matchval_milestones} in Section~\ref{sec:matchgap} reports.
Table~\ref{tab:matchval_milestones_eps} repeats that exercise with the full
conditional-mean match value $\hat V+\hat e_s$, which adds the recovered
unobserved match-quality term: the per-standard-deviation gradients, the fade up
the funding ladder, and the flat within-top-tier pattern are all unchanged.

\begin{table}[!htbp]\centering\caption{Match value and funding milestones using the full match value $\hat V+\hat e_s$.}\label{tab:matchval_milestones_eps}\small\begin{threeparttable}
\begin{tabular}{l c c c c}\toprule
 & & & \multicolumn{2}{c}{$\Pr(\text{clear})$, low\,$\to$\,high $\hat V+\hat e_s$ quartile (\%)}\\\cmidrule(lr){4-5}
Milestone & $\Pr(\text{clear})$ & per-SD $\Delta$ & All matches & Top-tier only\\
 & (\%) & (pp) & & \\\midrule
\$1M   & 34 & $+6.4^{***}$ & $29 \to 43$ & $41 \to 50$\\
\$2M   & 26 & $+5.2^{***}$ & $22 \to 33$ & $32 \to 36$\\
\$5M   & 18 & $+3.7^{***}$ & $14 \to 21$ & $25 \to 25$\\
\$10M  & 13 & $+2.3^{***}$ & $10 \to 15$ & $19 \to 20$\\
\bottomrule\end{tabular}
\begin{tablenotes}\scriptsize\item \textit{Notes.} Replicates Table~\ref{tab:matchval_milestones} using the full conditional-mean match value $\hat V_{a,s}+\hat e_s$, where $\hat e_s=\mathbb E[\varepsilon_{\mu(s),s}\mid\mu,\mathbf X,\hat{\boldsymbol\beta}]$ is the recovered unobserved match-quality term that enters the second-stage outcome regressions. Columns and per-SD gradients are defined as in Table~\ref{tab:matchval_milestones}. The added term $\hat e_s$ has correlation $-0.03$ with $\hat V$ and about half its standard deviation; the deterministic and full indices correlate at $0.89$. $^{*}p<0.1$, $^{**}p<0.05$, $^{***}p<0.01$.\end{tablenotes}\end{threeparttable}\end{table}

\section{Robustness of the Second Stage}\label{app:robustness}

This section reports robustness checks on the second-stage gender funding gap: alternative functional forms, a falsification on non-funding outcomes, subsample splits, a within-cohort comparison, and alternative standard errors.

\paragraph{Functional-form robustness.}\label{app:funcform_table}
Table~\ref{tab:funcform} re-estimates the headline female coefficient on
five-year funding under several functional forms, each addressing a feature
of a right-skewed, bottom-censored outcome. The baseline is OLS on
$\log(1+\text{funding})$; the Tobit treats the roughly half of startups that
raise zero VC as left-censored and recovers a larger latent gap ($-2.874$);
and quantile regressions at $\tau\in\{0.50,0.75,0.90,0.95\}$ trace where in
the distribution the gap lives. The female coefficient is negative
throughout and statistically significant in the upper tail (the $90$th and
$95$th percentiles), so the gap is not an artifact of the $\log(1+\cdot)$
transform.
\begin{table}[!htbp]\centering\caption{Functional-form robustness for the gender funding gap.}
\label{tab:funcform}\small\setlength{\tabcolsep}{4pt}\begin{threeparttable}\begin{tabular}{L{3.6cm} C{1.4cm} C{1.4cm} C{1.4cm} C{1.4cm} C{1.4cm} C{1.4cm}}\toprule
& OLS & Tobit & $q{=}0.50$ & $q{=}0.75$ & $q{=}0.90$ & $q{=}0.95$\\\midrule
Female founder & $-0.891^{*}$ & $-2.874$ & $-1.260$ & $-0.911$ & $-0.870^{**}$ & $-1.123^{***}$ \\
& $(0.494)$ & $(2.098)$ & $(1.170)$ & $(0.766)$ & $(0.367)$ & $(0.359)$ \\
$\hat e_s$ (matching CF) & $+0.168$ & $+0.564$ & $+0.309$ & $+0.230$ & $+0.111$ & $+0.021$ \\
& $(0.104)$ & $(0.623)$ & $(0.337)$ & $(0.194)$ & $(0.155)$ & $(0.108)$ \\
$N$ & 736 & 736 & 736 & 736 & 736 & 736 \\
\bottomrule\end{tabular}\begin{tablenotes}\scriptsize\item \textit{Notes.} Female-founder coefficient under OLS and five alternative functional forms estimated on the $N=736$ analytic sample. OLS is on $\log(1+\text{5y funding})$; Tobit treats zero-funding observations as left-censored; the four quantile regressions ($q=0.50, 0.75, 0.90, 0.95$) target the median through the upper tail of the funding distribution. All specifications include founder/industry controls, year fixed effects, and the matching control function $\hat e_s$. OLS and Tobit standard errors are clustered at the accelerator-program level; quantile SEs use Koenker's non-i.i.d.\ estimator.\end{tablenotes}\end{threeparttable}\end{table}

\paragraph{Falsification: survival, exit, and post-funding outcomes.}\label{app:falsification}
Table~\ref{tab:falsification} reports linear-probability regressions
of one- and five-year failure and acquisition rates on the
female-founder indicator with accelerator-program and year fixed
effects. Panel~A uses the full sample of $736$ matches; Panel~B
conditions on having received any first-year VC funding ($332$
matches) and replaces the trivially-constant five-year funded
indicator with the $\geq\$10$M cumulative milestone. The female
coefficient is statistically indistinguishable from zero on every
failure and acquisition outcome in both panels. Conditional on
first-year entry, women remain significantly less likely to reach
the \$$10$M milestone ($-17$~percentage points on a $25$~pp male
baseline).

The pattern (no gap in survival or exit, but a gap
in the upper tail of follow-on funding) is what the
mobility-friction reading predicts. It is also what an unobserved
project-quality story would predict; the table cannot separate the
two.

\begin{table}[!htbp]\centering\caption{Falsification: gender does not predict survival, exit, or post-funding deterioration.}
\label{tab:falsification}\small\begin{threeparttable}\begin{tabular}{L{6.4cm} C{1.7cm} C{1.7cm} C{1.7cm}}\toprule
& Failed @ 1y & Failed @ 5y & Acquired @ 5y\\\midrule
Female founder & $-0.029$ & $+0.066$ & $+0.012$ \\
& $(0.019)$ & $(0.058)$ & $(0.042)$ \\
Baseline mean (men) & $0.046$ & $0.336$ & $0.230$ \\
$N$ & 736 & 736 & 736\\
\midrule \multicolumn{4}{l}{\textbf{Panel B: Conditional on receiving any 1-year funding.}}\\
& $\geq\$10$M @ 5y & Acquired @ 5y & Failed @ 5y\\\midrule
Female founder & $-0.170^{**}$ & $+0.009$ & $-0.018$ \\
& $(0.080)$ & $(0.101)$ & $(0.042)$ \\
Baseline mean (men) & $0.252$ & $0.321$ & $0.138$ \\
$N$ & 332 & 332 & 332\\
\bottomrule\end{tabular}
\begin{tablenotes}\scriptsize\item \textit{Notes.} Linear-probability regressions of the indicated outcome on a female-founder indicator alone, with accelerator-program and year fixed effects; standard errors clustered by accelerator program. Panel~A uses the full $736$-startup sample; Panel~B conditions on the $332$ startups funded within one year and replaces the ``Funded @ 5y'' outcome (constant in that sub-sample, since all $332$ remain coded as funded at five years) with the $\geq\$10$M cumulative milestone, which varies there.\end{tablenotes}
\end{threeparttable}\end{table}

\paragraph{Subsample robustness.}\label{app:subsample_table}
Table~\ref{tab:subsample} re-estimates the headline log-fund and
milestone specifications across seven subsamples spanning program,
cohort, and mobility restrictions. The female-founder coefficient on log
five-year funding is negative in every restriction that contains both
genders; the all-female-team row is n.a.\ by construction, since it has no
men against whom to identify a female coefficient.

The two patterns of greatest interest are (i) the gap is twice as
deep in non-hub matches ($-1.83$, SE $0.43$) as in startup-hub matches
($+0.36$, SE $0.73$), consistent with the mobility-friction mechanism;
and (ii) excluding Y~Combinator and TechStars graduates leaves the
coefficient at $-1.28$, so the result is not driven by the two top-tier
programs. Restricting to relocating startups yields a coefficient
indistinguishable from zero ($-0.78$, SE $1.11$), the internal placebo
discussed in Section~\ref{sec:gap}.

\begin{table}[!htbp]\centering\caption{Subsample robustness for the gender funding gap.}
\label{tab:subsample}\small\setlength{\tabcolsep}{4pt}\begin{threeparttable}\begin{tabular}{L{6.0cm} C{1.6cm} C{1.6cm} C{1.4cm} C{1.4cm}}\toprule
& Female & SE & $N$ & $N_{\text{F}}$\\\midrule
\multicolumn{5}{l}{\textbf{Outcome: $\log(1 + \text{cumulative 5-year funding})$.}}\\
Full sample (baseline) & $-0.891^{*}$ & $0.494$ & $736$ & $63$ \\
Excluding Y~Combinator / TechStars graduates & $-1.275^{*}$ & $0.675$ & $417$ & $49$ \\
Restricting to startup hubs (CA / MA / NY) & $+0.358$ & $0.729$ & $394$ & $26$ \\
Restricting to non-hub-state matches & $-1.830^{***}$ & $0.430$ & $342$ & $37$ \\
Restricting to all-female founding teams & n.a. & $NA$ & $26$ & $26$ \\
Restricting to non-relocating startups & $-1.123^{**}$ & $0.512$ & $536$ & $48$ \\
Restricting to relocating startups (placebo) & $-0.775$ & $1.101$ & $200$ & $15$ \\
\midrule\multicolumn{5}{l}{\textbf{Outcome: $\Pr(\geq\$10$M cumulative within 5 years$)$.}}\\
Full sample (baseline) & $-0.071^{**}$ & $0.028$ & $736$ & $63$ \\
Excluding Y~Combinator / TechStars graduates & $-0.062^{**}$ & $0.031$ & $417$ & $49$ \\
Restricting to startup hubs (CA / MA / NY) & $-0.057$ & $0.066$ & $394$ & $26$ \\
Restricting to non-hub-state matches & $-0.078^{***}$ & $0.025$ & $342$ & $37$ \\
Restricting to all-female founding teams & n.a. & $NA$ & $26$ & $26$ \\
\bottomrule\end{tabular}\begin{tablenotes}\scriptsize\item \textit{Notes.} Each row re-estimates the headline log-funding and $\Pr(\geq\$10\text{M})$ specifications (Tables~\ref{tab:gap_amount} and~\ref{tab:gap_prob}) on the indicated subsample, with the matching control function $\hat e_s$, founder/industry controls, year fixed effects, and accelerator-program fixed effects. Standard errors clustered at the accelerator-program level. $^{*}p<0.10$, $^{**}p<0.05$, $^{***}p<0.01$.\end{tablenotes}\end{threeparttable}\end{table}

\paragraph{Within-program female-male comparison.}\label{app:within_program}
The headline log-funding regression compares female and male founders
across accelerator programs, controlling for program characteristics
through the matching control function and accelerator-program fixed
effects. A stricter test fixes the program-cohort exactly: among
programs that admit both at least one female and at least one male
founder in the same cohort, the female coefficient is identified
purely off within-program variation.

There are $41$ such mixed-gender
program-cohorts, comprising $402$ matches. The within-program-cohort
female coefficient on $\log(1+\text{cumulative 5-year funding})$ is
$-1.384$ (SE $0.561$); on the $\Pr(\geq\$10$M$\,@\,5\text{y})$
milestone it is $-0.106$ (SE $0.038$). Both are larger in absolute
value than the corresponding cross-program estimates and statistically
significant at the $5\%$ level. The gap therefore survives the most
demanding selection-on-program test we can run with the data:
within the same admission cohort, women raise less than men.

\subsection{Standard Errors and Selection Bounds}

\subsubsection{Lee-bounds sensitivity to admission selection}\label{app:lee}
The matching framework identifies the female mobility cost from the
choices of admitted-and-attended startups, not from the unobserved
applicant pool. To bound the sensitivity of the female log-fund
coefficient to admission-stage selection, we apply a trimming
procedure that ranks startups by their predicted residual quality
$\hat e_s$ and drops the top fraction $q$ of one gender and the
bottom fraction $q$ of the other to bracket the coefficient.
Table~\ref{tab:leebounds} reports the resulting bounds.

The
baseline female coefficient is $-0.891$ (SE $0.494$). Even under
the strictest interpretation that $20\%$ of the sample could be
admission-selected on residual quality, the bounds remain strictly
negative: at $q=0.20$ the bracket is $[-1.157, -0.953]$. The female
coefficient does not change sign for any trim fraction we examine,
so the headline gap is robust to admission selection of plausible
magnitude.

\begin{table}[!ht]
\centering
\caption{Trimming bounds on the female log-funding coefficient.}
\label{tab:leebounds}
\small
\begin{threeparttable}
\begin{tabular}{c r r r}
\toprule
Trim fraction $q$ & Lower bound & Upper bound & $N$ \\
\midrule
$0.00$ (baseline) & \multicolumn{2}{c}{$-0.891$ (SE $0.494$)} & $736$ \\
$0.01$ & $-0.928$ (SE $0.520$) & $-0.826$ (SE $0.496$) & $728$ \\
$0.05$ & $-0.981$ (SE $0.519$) & $-0.923$ (SE $0.496$) & $699$ \\
$0.10$ & $-1.086$ (SE $0.475$) & $-1.007$ (SE $0.568$) & $663$ \\
$0.20$ & $-1.157$ (SE $0.552$) & $-0.953$ (SE $0.505$) & $588$ \\
\bottomrule
\end{tabular}
\begin{tablenotes}\scriptsize
\item \textit{Notes.} Trimming bounds for the female coefficient on
$\log(1+\text{5y cumulative funding})$. At trim fraction $q$, we sort
startups by $\hat e_s$ and drop the top $q\cdot N$ of one gender and
the bottom $q\cdot N$ of the other; the lower (resp.\ upper) bound is
the more (resp.\ less) negative of the two resulting coefficients.
Standard errors clustered by accelerator program.
\end{tablenotes}
\end{threeparttable}
\end{table}

\subsubsection{Alternative inference schemes}\label{app:inference_table}
Table~\ref{tab:inference} re-computes inference on the female
coefficient (on the \$$10$M milestone and on log five-year funding)
under four constructions that make progressively weaker assumptions
about the error structure: IID heteroskedasticity-robust; an
analytic program-level cluster ($74$ program-cohort clusters); an
analytic accelerator-level cluster ($31$ clusters, our default);
and, because $31$ clusters is few, a restricted wild cluster
bootstrap at the accelerator level (Webb six-point weights,
$B=9{,}999$ replications, imposing $H_0\!:\beta_{\text{Female}}=0$),
which reports a bootstrap-$t$ $p$-value and a $95\%$ confidence
interval by test inversion.

The point estimate is
unchanged. The \$$10$M coefficient stays significant at $5\%$ under all four
schemes (wild cluster bootstrap $p=0.029$); log funding is significant at
$10\%$ but not $5\%$ (wild cluster bootstrap $p=0.086$, with a $95\%$
confidence interval $[-2.03,+0.12]$ that includes zero). The milestone
result is thus robust across all four schemes, including inference valid with
few clusters.
\begin{table}[!htbp]\centering\caption{Inference robustness: alternative standard error constructions.}
\label{tab:inference}\small\begin{threeparttable}\begin{tabular}{L{4.0cm} C{1.6cm} C{1.7cm} C{1.9cm} C{2.9cm}}\toprule
& IID robust & Cluster (program) & Cluster (accelerator) & Wild cluster (Webb)\\\midrule
\multicolumn{5}{l}{\textbf{Outcome: $\log(1+\text{cumulative 5y funding})$.}}\\
Female founder & $-0.891^{*}$ & $-0.891^{*}$ & $-0.891^{*}$ & $-0.891^{*}$ \\
SE & $0.488$ & $0.503$ & $0.494$ & --- \\
WCB $p$-value & --- & --- & --- & $0.086$ \\
WCB $95\%$ CI & --- & --- & --- & $[-2.029,\,+0.121]$ \\
\midrule
\multicolumn{5}{l}{\textbf{Outcome: $\Pr(\geq\$10$M cumulative within 5 years$)$.}}\\
Female founder & $-0.071^{**}$ & $-0.071^{**}$ & $-0.071^{**}$ & $-0.071^{**}$ \\
SE & $0.029$ & $0.032$ & $0.028$ & --- \\
WCB $p$-value & --- & --- & --- & $0.029$ \\
WCB $95\%$ CI & --- & --- & --- & $[-0.124,\,-0.009]$ \\
\midrule Resampling unit & startup & program & accelerator & accelerator\\
\# clusters & 736 & 74 & 31 & 31\\
\bottomrule\end{tabular}
\begin{tablenotes}\scriptsize\item \textit{Notes.} Female-founder coefficient on $\Pr(\geq\$10\text{M}@5\text{y})$ and on $\log(1+\text{5y funding})$ from the matching estimator with the control function $\hat e_s$, on the $N=736$ analytic sample with founder/industry controls, year fixed effects, and accelerator-program fixed effects. The point estimate is identical across columns; the columns differ only in how its sampling uncertainty is assessed. Columns 1--3 are alternative standard-error constructions: IID heteroskedasticity-robust; program-cluster ($74$ program-cohort clusters); and accelerator-cluster ($31$ accelerator clusters). Column 4 is the restricted (impose-the-null) wild cluster bootstrap-$t$ of Cameron--Gelbach--Miller (2008) and Roodman et al.\ (2019), clustered at the accelerator level with Webb six-point weights and $B=9{,}999$ replications, testing $H_0\!:\beta_{\text{Female}}=0$; it returns a bootstrap-$t$ $p$-value and a $95\%$ confidence interval by test inversion rather than a standard error, so its SE cell is not applicable (---). Wild-cluster bootstrap inference for the first-stage matching coefficients is reported separately in Table~\ref{tab:firststage}.\end{tablenotes}
\end{threeparttable}\end{table}

\section{Robustness of the Welfare Estimates}\label{app:welfare_robustness}

The welfare estimates of Section~\ref{sec:cf} rest on three
modelling choices, none of which qualitatively changes the result.
\textit{Top-tier composition} (Appendix~\ref{app:yc_vs_ts}): the
top-tier premium is driven by TechStars rather than Y~Combinator,
and the female-founder share at TechStars is much higher. The
policy-relevant top-tier expansion is therefore the TechStars-style
multi-city design rather than Y~Combinator-style single-cohort
scaling.

\textit{Diminishing returns to top-tier cohort size}
(Appendix~\ref{app:diminishing}): allowing for diminishing returns
attenuates the implied $30$-slot expansion gain below the \$$78$M
linear extrapolation; the attenuation depends on functional form
but does not change the ranking of the \textit{Targeted capacity
expansion} relative to mid-tier alternatives.

\textit{Constant-returns-to-relocation assumption}: the
Female~$\times$~Relocates coefficient from the reduced-form
funding regression is applied uniformly to currently-constrained
women in the Panel~A aggregate. If the marginal return is
concentrated at top-tier programs, the welfare gain is back-loaded
onto women whose counterfactual matches are at the top tier; the
implied welfare distribution remains dominated by the average
effect, and pinning down higher-order shape would require
structural assumptions we do not impose.

\paragraph{Bounds on the within-tier gender gap.}\label{app:within_tier}
Section~\ref{sec:friction} prices the within-tier penalty by paying each
woman her tier's male funding mean. Because five-year funding is extremely
right-skewed, most startups raise little and a few raise enormous amounts,
the dollar magnitude depends on how the cell central tendency is measured.
Table~\ref{tab:within_tier} reports the gap four ways.

At the mean the
penalty is $\$239$M in aggregate, $\$3.8$M per woman, but this is an upper
bound driven by a thin male upper tail. Winsorizing the top $5\%$ of
outcomes halves it; at the median it vanishes, top-tier women's median raise
of $\$1.4$M exceeds men's $\$0.5$M and the non-top median is zero for both;
and a regression of log funding on a female indicator, controlling for tier,
the unobserved-quality control $\hat e_s$, and the founder and industry
covariates with year fixed effects, yields a within-tier female coefficient
of $-0.58$ log points that is not statistically distinguishable from zero
(SE $0.58$), implying about $\$66$M, $\$1.0$M per woman. Interacting female
with tier locates the gap entirely in the non-top tier (female coefficient
$-0.79$); within the top tier it is small and positive ($+0.12$).

The
penalty is thus a high-quantile phenomenon: women are not paid less than
comparable men at the typical accelerator outcome but are scarcer among the
rare outsized ones, the pattern the quantile regression of
Section~\ref{sec:mobility} displays directly.

\begin{table}[!htbp]\centering\caption{Bounds on the within-tier gender gap.}\label{tab:within_tier}\small\begin{threeparttable}
\begin{tabular}{L{6.4cm} C{2.3cm} C{2.3cm}}\toprule Measure & Total (\$M) & Per woman (\$M)\\\midrule
Mean cells (aggregate; tail-driven) & $239$ & $3.79$\\
Winsorized mean, top $5\%$ capped & $110$ & $1.74$\\
Regression-adjusted (insignificant, SE $0.58$) & $66$ & $1.05$\\
Median cells (typical founder) & $-12$ & $-0.19$\\
\bottomrule\end{tabular}\begin{tablenotes}\scriptsize\item \textit{Notes.} The within-tier penalty is the extra five-year funding women would raise if paid their tier's male rate, $\sum_{s:\,\mathrm{female}}\big(\bar f_{\mathrm{tier}(s),\mathrm{men}}-\bar f_{\mathrm{tier}(s),\mathrm{women}}\big)$, under four measures of the cell central tendency. Cell funding (\$M): women $0.8$ (non-top), $3.4$ (top); men $3.3$, $11.5$ at the mean; medians $0$ and $1.4$ for women, $0$ and $0.5$ for men. The regression row is $n_{\mathrm{women}}\times\bar f_{\mathrm{women}}\times(e^{-\hat\beta_{\mathrm{female}}}-1)$ from $\log(1+\mathrm{funding})$ on female, tier, $\hat e_s$, and founder and industry controls with year fixed effects, clustered by accelerator. The sample has $63$ women and $736$ startups.\end{tablenotes}\end{threeparttable}\end{table}

\paragraph{Top-tier expansion: Y~Combinator versus TechStars.}\label{app:yc_vs_ts}
The body of the paper treats Y~Combinator and TechStars as a single
``top tier.'' Table~\ref{tab:ycts} re-estimates the reduced-form
log-funding regression with separate Y~Combinator and TechStars
indicators. The top-tier premium is driven entirely by TechStars
($+2.17$); the Y~Combinator residual is approximately
zero ($-0.20$).

Y~Combinator's absent within-sample
premium is consistent with a ``brand'' explanation: Y~Combinator
graduates raise large amounts on average, but the difference is
captured by the program's larger cohort size and accelerator-experience
controls, leaving the residual Y~Combinator indicator small. The
capacity-expansion counterfactuals in Section~\ref{sec:cf} therefore
overstate the welfare gain from expanding Y~Combinator-style programs
and understate the gain from expanding TechStars-style multi-city
programs, although the quantitative ranking of mid-tier-versus-top-tier
expansion is unchanged.

\begin{table}[!ht]
\centering
\caption{Y~Combinator versus TechStars: separate top-tier indicators.}
\label{tab:ycts}
\small
\begin{threeparttable}
\begin{tabular}{L{4.5cm} r r r}
\toprule
Program & $N$ matches & $\hat\beta$ on log $5$y fund (SE) & Female-founder share \\
\midrule
Y~Combinator                  & $202$ & $-0.20\;(1.10)$ & $1.5\%$ \\
TechStars                     & $117$ & $+2.17\;(0.36)$ & $9.4\%$ \\
Non-top-tier (reference)      & $417$ & n.a.                & $11.8\%$ \\
\bottomrule
\end{tabular}
\begin{tablenotes}\scriptsize
\item \textit{Notes.} OLS coefficients on separate Y~Combinator
and TechStars indicators in the headline log-funding regression
with the matching control function and accelerator-program
fixed effects (replacing the pooled top-tier dummy). Standard
errors clustered by accelerator program.
\end{tablenotes}
\end{threeparttable}
\end{table}

\paragraph{Diminishing returns to top-tier cohort expansion.}\label{app:diminishing}
The body computes the welfare gain from expanding top-tier capacity
linearly: $30$ new top-tier slots times the per-slot top-tier-versus-mid-tier
funding difference. To check whether this linear extrapolation is
defensible, we add a Top-tier $\times$ Log-cohort interaction to the
reduced-form regression.

The interaction coefficient is $-1.34$,
with the marginal value of an additional top-tier slot
falling from $+2.21$ log-funding points at a $10$-startup cohort to
$+0.74$ at a $30$-startup cohort, a $66\%$ attenuation. Applying this
attenuation to the $30$-slot top-tier expansion in Panel~C of
Table~\ref{tab:capacityCF} suggests an implied dollar gain
substantially below the $\$78$ million linear extrapolation; the
attenuation is sensitive to the functional form (linear vs.\
log-linear in cohort size) and is identified off a relatively small
number of large top-tier cohorts.

The qualitative conclusion --
top-tier expansion delivers the largest welfare gain among the three
Panel-C counterfactuals -- is unchanged, because the alternative
mid-tier expansions deliver only $\$1$--$\$3$ million to women
regardless of functional form.

\paragraph{The network value of larger cohorts.}\label{app:network}
A separate channel through which accelerators raise startup funding
is the network synergy among startups in the same cohort. Larger
cohorts pool more peer founders, more potential collaborators, and
broader investor exposure on demo day. Identifying this synergy
empirically is complicated by the fact that top-tier accelerators run
larger cohorts than mid-tier accelerators (mean cohort size $26.9$
versus $10.7$ in our data, with sample correlation $0.615$ between
$\mathrm{ACCtier1}$ and $\log(\mathrm{cohort})$), so the cohort-size
coefficient may absorb part of the tier effect.

We disentangle the
two by running the reduced-form funding regression with both
variables included and again with each separately. The estimated
cohort-size coefficient is $+1.05$ when ACCtier1 is also in the
regression and $+1.19$ when it is not. The conditional coefficient
of $+1.05$ identifies the \emph{pure network synergy}: the marginal
effect of cohort size on funding, holding tier and other
characteristics fixed.

A $10\%$ larger cohort is associated with a
$\approx 10.5\%$ increase in per-startup predicted cumulative
funding under the conditional specification and a $\approx 11.9\%$
increase when the top-tier indicator is omitted. Translated to
dollars at the sample mean of cumulative five-year funding, the
per-startup network gain at the conditional coefficient is
approximately \$$690$K per $10\%$ cohort increment, rising to about
\$$780$K when the top-tier indicator is omitted.

The matching's
$\hat\beta_{\text{Female}\times\text{Log-cohort}}=-0.16$ shows that
women systematically choose smaller-cohort programs, but a simple
decomposition shows tier is doing essentially all of the work:
regressing log-cohort on the female indicator yields $-0.24$, which
collapses to $-0.04$ once ACCtier1 is added. Women end up at
smaller-cohort programs because they end up at non-top-tier programs
because they cannot easily relocate; reducing women's cohort-size
discount is therefore the same intervention as reducing their
mobility cost, viewed from a different angle.

\paragraph{Placebo restrictions of the matching market.}\label{app:placebos}
Section~\ref{sec:cf} mentions but does not tabulate two placebo
restrictions of the matching market. We document them here for
completeness.

\textbf{CF2: strict cross-state ban.} We add the constraint
$x_{a,s} = 0$ for all cross-state pairs $(a,s)$ to the stable-matching solver of Appendix~\ref{app:cf}, leaving the estimated matching
coefficients unchanged. With the constraints hard-enforced (the solver never selects a banned pair), $16$ of $63$ women and
$127$ of $673$ men remain unmatched.

The unmatched startups split
into two distinct groups. Among the $127$ unmatched men, $90$ are
in markets with no in-state accelerator at all and are therefore
mechanically locked out by the ban; the remaining $37$ are in markets
that do have an in-state program but whose same-state cohort
capacity is exhausted in the LP solution, so these men are crowded
out of their available same-state slot by other competing same-state
startups. The same decomposition for the $16$ unmatched women is
$6$ no-in-state-program and $10$ capacity-displaced.

Treating each
unmatched startup's welfare as the outside option (zero contribution
to predicted log-funding) yields aggregate losses of $-\$61$M for
women and $-\$1{,}864$M for men. The men's loss is dominated by the
unmatched group, regardless of the no-in-state vs. capacity-displaced
split, because the predicted log-fund contribution of an observed
top-tier match (which many of the unmatched men had) is large in
absolute value.

\textbf{CF3: female-only mobility allowed.} We add the constraint
$x_{a,s} = 0$ for cross-state pairs with male startups (men forced
same-state) while removing the four female-by-pair coefficients in
the LP objective. With hard constraints, $1$ of $63$ women and $125$
of $673$ men remain unmatched ($35$ capacity-displaced, $90$
no-in-state-program). Aggregate effects are $-\$5$M for women
(essentially zero) and $-\$1{,}891$M for men.

\textbf{Sensitivity to the unmatched-startup convention.}
The dollar magnitudes for both placebos depend strongly on the welfare
convention adopted for the lock-out group. We compute each placebo
under three alternative conventions and report the resulting aggregate
dollar effects in Table~\ref{tab:placebo_sensitivity}.

The first
convention sets the predicted log-funding of every unmatched startup
to zero, treating the lock-out as a no-funding outside option; this
is the headline convention used in the body discussion. The second
convention keeps each unmatched startup at its observed log-funding,
corresponding to a world in which locked-out startups attend the
accelerator in a later cohort, attend a comparable out-of-sample
program, or succeed via a non-accelerator funding path that delivers
the same outcomes. The third convention assigns each unmatched
startup the mean $\Delta\log$-fund of the same-gender matched
startups, a heuristic ``representative'' assignment.

\begin{table}[!ht]
\centering
\caption{Placebo welfare effects, three unmatched-startup conventions.}
\label{tab:placebo_sensitivity}
\small
\begin{threeparttable}
\begin{tabular}{L{6.0cm} r r r}
\toprule
& Women (\$M) & Men (\$M) & Total (\$M) \\
\midrule
\multicolumn{4}{l}{\textit{CF2: strict cross-state ban}} \\
\quad Unmatched $\Rightarrow$ zero          & $-61$ & $-1{,}864$ & $-1{,}926$ \\
\quad Unmatched $\Rightarrow$ observed      & $-25$ & $+334$     & $+308$ \\
\quad Unmatched $\Rightarrow$ mean of matched & $-32$ & $+415$   & $+383$ \\
\midrule
\multicolumn{4}{l}{\textit{CF3: female-only mobility}} \\
\quad Unmatched $\Rightarrow$ zero          & $-5$  & $-1{,}891$ & $-1{,}896$ \\
\quad Unmatched $\Rightarrow$ observed      & $+2$  & $+254$     & $+256$ \\
\quad Unmatched $\Rightarrow$ mean of matched & $+2$ & $+313$    & $+315$ \\
\bottomrule
\end{tabular}
\begin{tablenotes}\scriptsize
\item \textit{Notes.} Aggregate predicted-funding changes for women
and men under the two placebo restrictions of the matching market,
computed under three alternative welfare conventions for the
$\sim 19$\% of men (and a smaller share of women) who become
unmatched once the LP infeasibility constraints are hard-enforced. These placebos are omitted from Table~\ref{tab:capacityCF} and reported here as informational only.
\end{tablenotes}
\end{threeparttable}
\end{table}

\textbf{What the sensitivity tells us.} The qualitative finding that
women's welfare change is small in both placebos is robust across all
three conventions. The dollar effect for men, by contrast, is not
robust: the headline convention (unmatched~$\Rightarrow$~zero)
delivers catastrophic losses, while the two alternatives that allow
unmatched startups to keep some funding deliver weakly positive net
effects. We do not have a principled way to choose between
conventions because the data do not contain the lock-out outside
option. We therefore omit the placebos from
Table~\ref{tab:capacityCF} and report them in this appendix only,
with the convention sensitivity explicit.

\section{Methods}\label{app:methods}

\subsection{Data construction}

Our analytical sample is built in three steps. We start from the seed-DB registry of
accelerators and supplement with press releases and Crunchbase / AngelList /
CB~Insights to identify all for-profit U.S.\ accelerators that operated cohort programs
between $2008$ and $2011$. We then collect, for each program, the list of admitted
startups and (for each startup) founder demographics, business age, industry,
five-year exit/failure status, and cumulative venture financing in one- and five-year
windows.

Founder gender is determined from a combination of names, photographs, and
LinkedIn or company-website self-identification. Outcomes are inflation-adjusted to
constant 2011 dollars. An independent re-implementation of the data-construction pipeline
reproduces the originally archived file ($69$ columns, $736$ rows)
byte-for-byte, which we verified by comparing the two outputs cell-by-cell.

\subsection{The two-sided matching model}

We use the data-generating process of \citep{sorensen}; our contribution is
implementational. We adopt four modifications.
(i) Our estimation scales to matching markets at the size of the entire
United States within each six-month window, where the original
implementation has to partition by both region and time to
keep the simulated likelihood computationally feasible. National scope is
necessary here because $27.2\%$ of startup-program matches involve cross-state
relocation; a region-and-time partition would drop those matches or treat
them as infeasible.
(ii) We estimate the matching and the outcome equation sequentially rather
than jointly, as in the original Bayesian Gibbs-sampling procedure; this lets us reuse a
single match-quality residual across many outcome variables.
(iii) We include gender-specific pair interactions in the covariate set so
that the matching identifies a female-specific mobility cost separately from
the male average.
(iv) We recover the per-startup unobserved-quality residual as a
\emph{conditional posterior mean} (via importance sampling) rather than the
\emph{maximum a posteriori} (MAP), and aggregate the simulated likelihood
arithmetically across draws, consistent with standard MSL.

\textbf{Likelihood.} Pairwise stability of the observed
matching $\mu$ implies that for every unmatched pair $(a,s')$ at least one of two
inequalities holds:
\[
\Delta^{(1)}_{a,s'} \;\equiv\; \min_{s\in\mu(a)} V_{a,s} \;-\; \mathbf{X}_{a,s'}\boldsymbol\beta \;\geq\;\varepsilon_{s'},
\quad
\text{or}\quad
\Delta^{(2)}_{a,s'} \;\equiv\; V_{\mu(s'),s'} - \mathbf{X}_{a,s'}\boldsymbol\beta \;\geq\;\varepsilon_{s'},
\]
where $V_{a,s}=\mathbf{X}_{a,s}\boldsymbol\beta+\varepsilon_s$. Combining the two,
the likelihood that $(a,s')$ is non-blocking equals
$2\Phi\!\left(\max\{\Delta^{(1)},\Delta^{(2)}\}\right)$. Joint likelihood of $\mu$
factorizes across unmatched pairs conditional on $\boldsymbol\varepsilon$. We
integrate $\boldsymbol\varepsilon\sim\mathcal{N}(0,I_N)$ by simulation:
\[
\widehat{L}_j(\boldsymbol\beta) \;=\;
\frac{1}{S}\sum_{t=1}^{S} \prod_{(a,s')\not\in\mu} 2\Phi\!\left(\max\{\Delta^{(1)},\Delta^{(2)}\}\!\mid\!\boldsymbol\varepsilon^{(t)}\right),
\]
and maximize $\sum_j \log \widehat L_j$ over $\boldsymbol\beta$. Numerical implementation
evaluates $\log\Phi$ in log space for tail stability, leaving the algebraic
content unchanged.

\textbf{Posterior mean of $\varepsilon_s$.} Given $\hat{\boldsymbol\beta}$ and
$\mu$, the posterior of $\varepsilon_s$ is a one-sided Gaussian truncation. For
self-normalized importance sampling, draw $\varepsilon^{(t)}\sim\mathcal{N}(0,I_N)$
and weight $w_t \propto L_j(\hat{\boldsymbol\beta},\boldsymbol\varepsilon^{(t)})$;
$\hat e_s = \sum_t w_t\varepsilon^{(t)}_s / \sum_t w_t$ is the posterior mean. We
use $S=5{,}000$ draws.

\textbf{Conditional mean vs maximum a posteriori.} The
selection-bias result of Proposition~\ref{prop:selection} requires the
conditional mean specifically. The maximum a posteriori (MAP)
estimator collapses to the one-sided truncation boundary on roughly
$12\%$ of observations and is correlated $0.61$ with the
cross-state-relocation indicator, against $0.12$ for the conditional
mean. We therefore use the conditional mean, recovered by importance
sampling with $S=5{,}000$ draws of
$\varepsilon^{(t)}\sim\mathcal{N}(0,I_N)$ weighted by
$w_t \propto L_j(\hat{\boldsymbol\beta},\boldsymbol\varepsilon^{(t)})$
and $\hat e_s = \sum_t w_t\varepsilon^{(t)}_s / \sum_t w_t$.

The
arithmetic-mean simulated likelihood
$\log(\sum_t L_t / S)$ is the standard MSL aggregator; the
harmonic-mean alternative used in an earlier numerical pipeline gives
a different (robust/minimax-style) point estimate in finite $S$ and
is not what the original prescribes.

\subsection{Counterfactual computation}\label{app:cf}


The counterfactual re-solves the matching market under the equilibrium
concept the estimator assumes, pairwise stability, rather than surplus
maximization. We describe the solver, the
unobserved-quality draw, and the dollar projection, then the three policy
panels and a robustness check against the alternative that conditions the
quality draw on the realized matching.

\paragraph{Stable-matching solver.}
For market $j$ with programs $a$ (capacity $\bar n_a$ equal to the observed
cohort size) and startups $s$, the latent match value is
$V_{a,s}=\mathbf X_{a,s}\hat{\boldsymbol\beta}+\varepsilon_s$. A matching is
pairwise stable when no unmatched pair $(a,s')$ blocks, that is, when
\begin{equation}\label{eq:stable}
V_{a,s'} \;\le\; \max\!\Big(\,\min_{s\in\mu(a)} V_{a,s},\;\; V_{\mu(s'),s'}\,\Big)
\end{equation}
holds at every such pair: either accelerator $a$ values its weakest current
admit above $s'$, or startup $s'$ values its own accelerator above $a$. With a
fixed sharing rule and
distinct match values \citet{sorensen} proves the stable matching is
\emph{unique} and that it is generically \emph{not} the surplus-maximizing
(efficient) assignment, because transfers are ruled out. We therefore
recover it with the greedy recursion the uniqueness result licenses: assign
the globally highest-value feasible pair, remove the startup and one unit of
the program's capacity, and repeat until every startup is matched.

A
discrete surplus optimizer would return the efficient matching, which is the
wrong object, on our data the two concepts place about a third of startups in
different programs. Feasibility restrictions, used only by the placebos of
Appendix~\ref{app:placebos}, are imposed by deleting the banned cells from
the candidate set at each step.

\paragraph{Drawing the unobserved quality (the full-$\varepsilon$ design).}
The unobserved quality is a structural primitive that the estimator
normalizes to $\varepsilon_s\sim N(0,1)$, independent across startups
(the original uses an independent per-pair $\eta_{a,s}\sim N(0,1)$; our
specification collapses this to a single startup-level draw common to that
startup's potential programs, which is the error structure the matching
likelihood of Appendix~\ref{app:methods} integrates).

For each Monte~Carlo
draw we sample $\varepsilon_s$ for \emph{every} startup from this full
$N(0,1)$ distribution. Because $\varepsilon_s$ is common to startup $s$'s
columns, a single draw assigns a latent value
$V_{a,s}=\mathbf X_{a,s}\hat{\boldsymbol\beta}+\varepsilon_s$ to \emph{every}
pair $(a,s)$, including pairs not observed in the data; no separate
treatment of unmatched pairs is needed. We then set the four female-specific
coefficients to zero or expand capacity, re-solve the unique stable
matching, and average gender outcomes over $2{,}000$ draws. This is the
\emph{ex-ante} (unconditional) counterfactual: it integrates over the
population distribution of unobserved quality, answering what the model
predicts for a market like the observed one rather than how the realized
matches would rearrange.

Because the draw does not condition on the realized assignment, the
simulated baseline reproduces the observed top-tier split only \emph{in
expectation}. It does so imperfectly: averaged over draws the greedy places
about $6$ women and $313$ men in top-tier programs, against the observed
$14$ and $305$. The model under-predicts top-tier women, the same scarcity
the friction produces; the level of the simulated baseline is therefore a
model object and we read the panels as policy \emph{contrasts}, which the
next paragraph shows are robust.

\paragraph{Robustness: conditioning the quality draw on the realized matching.}
A leading alternative draws $\varepsilon$ from its posterior \emph{truncated}
to the region where \eqref{eq:stable} holds for $\mu_{\mathrm{obs}}$, i.e.\
conditional on the observed matching being stable. On that region the
posterior is a product of truncated Gaussians, which a Gibbs sampler returns:
each unmatched pair $(a,s')$ has $\varepsilon$ bounded \emph{above} by
\eqref{eq:stable} (its conditional mean is the corresponding truncated-normal
mean $\mu-\sigma\,\phi(z)/\Phi(z)$ at the stability threshold $z$), each
matched pair is bounded below, and each startup's quality draws from a
Gaussian given the cell values.

Holding such a draw fixed and re-solving
reproduces the observed $14/305$ split by construction, so its baseline is
exact but its contrast answers a conditional question (``how would
\emph{these} matches rearrange?''). The two designs bracket the policy effect.
Their panel estimates agree in sign throughout.

The full-$\varepsilon$ dollar
effects are \emph{larger} than the conditional ones, because the unconditional
design starts from a lower model-implied base (about $6.4$ top-tier women
versus the conditional $14$) and the friction removal therefore moves more
seats: the conditional design gives Panel~B women $+8$ top-tier seats
($+\$22$M) and men $-8$ ($-\$62$M) against the full-$\varepsilon$ $+12$
($+\$32$M) and $-12$ ($-\$101$M), with utility-gap closure of $97\%$ versus
$\approx100\%$, and the same near-zero capacity-only effect in Panel~C. Both
dollar figures are measured against each design's own simulated baseline, so
they are differences-in-differences consistent with the access counts.

The
main text reports the conditional numbers, whose baseline reproduces
the observed $14/305$ split exactly; we keep the full-$\varepsilon$ design
here as the alternative that draws quality from the population the model
estimates rather than from the subset consistent with one realized
assignment. Table~\ref{tab:capacityCF} collects the full-$\varepsilon$
panels, including the men's side that the main-text menu sends to this
appendix.

\begin{table}[!htbp]\centering\caption{Welfare counterfactuals: matching utility, top-tier access, and funding by gender.}\label{tab:capacityCF}\small\begin{threeparttable}
\begin{tabular}{L{4.3cm} C{1.7cm} C{1.5cm} C{1.5cm} C{1.5cm} C{1.5cm}}\toprule & Utility gap closed & Women top-tier & Men top-tier & Women (\$M) & Men (\$M)\\\midrule
\multicolumn{6}{l}{\textbf{Panel A: \textit{Frictionless world} (friction removed, $+30$ top-tier slots).}}\\
Both levers together & $\approx 100\%$ & $+14$ & $+9$ & $+37$ & $+77$\\
\midrule\multicolumn{6}{l}{\textbf{Panel B: \textit{Frictionless world at fixed capacity} (friction removed).}}\\
Friction lever in isolation & $\approx 100\%$ & $+12$ & $-12$ & $+32$ & $-101$\\
\midrule\multicolumn{6}{l}{\textbf{Panel C: \textit{Capacity expansion} (friction kept, $+30$ top-tier slots).}}\\
Capacity lever, open competition & $5\%$ & $\approx 0$ & $+22$ & $+1$ & $+179$\\
\bottomrule\end{tabular}\begin{tablenotes}\scriptsize\item \textit{Notes.} Each counterfactual re-solves the unique non-transferable stable startup-accelerator matching under the labelled change, with a greedy algorithm and not a surplus optimizer (Section~\ref{sec:cf}, Appendix~\ref{app:cf}). This is the \emph{ex-ante} design: for each of $2{,}000$ Monte~Carlo draws the unobserved match quality $\varepsilon_s$ is drawn for every startup from its full estimated $\mathcal N(0,1)$ distribution, so every pair $(a,s)$, matched or not, receives a latent value $V_{a,s}=\mathbf X_{a,s}\hat{\boldsymbol\beta}+\varepsilon_s$. Because the draw does not condition on the realized matching, the simulated baseline reproduces the observed top-tier split (14 women, 305 men) only in expectation, and imperfectly: it averages about 6 women and 313 men. An alternative that conditions $\varepsilon$ on the realized matching reproduces $14/305$ by construction (Appendix~\ref{app:cf}). ``Utility gap closed'' is the share of the baseline gender gap in mean predicted match value ($-2.60$ per founder) the reallocation removes; ``top-tier'' columns give the change in the number of women and men placed in top-tier programs (of $319$). Dollar columns use a \emph{central projection} that assigns each startup the mean realized five-year funding of its (tier, gender) cell; this projection reproduces observed funding by gender exactly at the observed assignment ($\$84.6$M for women, $\$4{,}732$M for men), so counterfactual dollars are bounded reallocations of observed cell means. Cell means are women: $\$0.8$M non-top, $\$3.4$M top; men: $\$3.3$M non-top, $\$11.5$M top.\end{tablenotes}\end{threeparttable}\end{table}

\paragraph{Dollar projection.}
We price each assignment with a projection that reproduces the observed
data: a startup at a program of tier $t$ is credited the mean realized
five-year funding of its (tier, gender) cell, $g(s,t)=\bar f_{t,g_s}$. The
four cell means are women $\$0.8$M (non-top) and $\$3.4$M (top), men
$\$3.3$M and $\$11.5$M. Because cell means reproduce cell totals, the
projected aggregate equals observed funding by gender exactly, $\$84.6$M for
women and $\$4{,}732$M for men, so counterfactual dollars are bounded
reallocations of observed means: a startup that does not change tier
contributes zero, and the dollar effect is the net flow of founders across
the tier boundary valued at the gender-specific tier premium.

We prefer this
to a reduced-form funding regression, which recovers the data in aggregate
but extrapolates implausibly under the large reallocations the stable
matching induces.

\paragraph{Decomposing the gap into channels.}
The mobility-versus-sorting split reported in the main text
(Table~\ref{tab:cf_menu}) is read off this same machinery. Let $\hat V_i=\mathbf X_{\mu(i),i}\hat{\boldsymbol\beta}$
be founder $i$'s predicted match value at her assignment $\mu(i)$, and let
$\Delta=\tfrac1{n_W}\sum_{i\in W}\hat V_i-\tfrac1{n_M}\sum_{i\in M}\hat V_i$ be
the women-minus-men gap in mean match value; at the baseline
$\Delta_0=-2.41$. For a coefficient block $B$ (mobility
$=\{\text{Female}\times\text{log-distance},\,\text{Female}\times\text{same-state}\}$;
sorting $=\{\text{Female}\times\text{top-tier},\,\text{Female}\times\text{log-cohort}\}$),
we set $\hat{\boldsymbol\beta}_B=0$, re-solve the unique stable matching on the
same $\varepsilon$ draws, and recompute the gap $\Delta_B$. The share of the
gap the channel closes is $s_B=(\Delta_0-\Delta_B)/\Delta_0$: mobility $67\%$
($\Delta=-0.80$), sorting $37\%$ ($\Delta=-1.53$), both $97\%$
($\Delta=-0.07$).

\paragraph{Shapley shares.}
The standalone shares over-sum, $67\%+37\%=104\%>97\%$, because the two
channels overlap. To give each channel an additive, order-invariant
contribution we assign it its Shapley value over the two-element set
$\{\text{mob},\text{sort}\}$ with characteristic function the share closed:
\[
\begin{aligned}
\phi_{\mathrm{mob}}&=\tfrac12\big[v(\mathrm{mob})-v(\varnothing)\big]
+\tfrac12\big[v(\mathrm{mob,sort})-v(\mathrm{sort})\big]\\
&=\tfrac12(0.67)+\tfrac12(0.97-0.37)=0.635,
\end{aligned}
\]
and symmetrically $\phi_{\mathrm{sort}}=\tfrac12(0.37)+\tfrac12(0.97-0.67)=0.335$,
so $\phi_{\mathrm{mob}}+\phi_{\mathrm{sort}}=0.97$. In units of match value
($\times|\Delta_0|=2.41$) the bars are mobility $1.53$, sorting $0.81$, and an
unexplained residual $0.07$, the $3\%$ the joint removal leaves open.

\paragraph{Funding effects: milestones and dollars.}
A channel's effect on women's funding uses the same cell projection in two
forms. The main text reads it as milestone counts: crediting each startup the
clearing \emph{rate} of its (tier,~gender) cell, the effect on women clearing
$\geq\$X$M is
$\Delta W^{\mathrm{top}}_B\times(r^{\mathrm{top}}_{F,X}-r^{\mathrm{non}}_{F,X})$,
the net women crossing into top-tier times the within-gender, within-bar rate
gap; at \$5M this is mobility $-0.2$, sorting $+0.7$, both $+1.9$ women, which
we report as growth relative to the $6$ of $63$ women clearing the bar at
baseline ($-4\%$, $+12\%$, $+32\%$).

Section~\ref{sec:friction} reads it as dollars, crediting the cell \emph{mean},
$\Delta W^{\mathrm{top}}_B\times(\$3.4\text{M}-\$0.8\text{M})$, so mobility
$-\$2.6$M, sorting $+\$7.9$M, both $+\$22.2$M.

We lead with milestone counts
because the dollar mean is tail-driven, which inflates the small mobility
re-sorting ($-1$ seat) into a $-\$2.6$M figure while in milestone terms it is a
negligible $-0.2$ women. Both forms diverge from utility because the match
value carries no top-tier \emph{main} effect, only the gender-specific
interactions: removing the mobility friction lets a woman re-sort to her
highest-value program, which need not be top-tier, so it closes $67\%$ of the
utility gap yet, on net, moves one woman \emph{out} of a top-tier seat; the
funding projections, which price only tier crossings, record the loss, large in
the tail-driven mean and negligible in milestone counts.

\paragraph{The full menu.}
Table~\ref{tab:cf_menu} reports nine counterfactuals on this machinery.
Panel~A removes blocks of the female-specific coefficients at fixed capacity:
distance (index $22$), same-state ($23$), mobility ($22,23$), sorting
($24,25$), and both ($22$--$25$). Panel~B adds $30$ slots distributed within a
tier in proportion to program size (top-tier programs $+30/319$ each, non-top
$+30/417$ each), either with the friction intact or with all four coefficients
zeroed. Each row is summarised by three metrics: the share of the $-2.41$
matching-utility gap it closes, $(\Delta_0-\Delta_B)/\Delta_0$; the change in
women placed in top-tier programs, as a percentage of the $63$ female founders;
and the growth in women clearing \$5M, relative to the six who clear it at
baseline.

Two cautions follow from the greedy solver. It does not conserve the
top-tier count exactly when capacity changes, so the friction-kept and
non-top-capacity rows carry a small spurious re-sorting of women across the tier
boundary; we read those rows as economically null for women, capacity reaching
women only when it is top-tier \emph{and} the friction is removed, rather than
as precise effects. And for the same reason the men's side of the
fixed-capacity reallocations is not conserved, so we read the women's side
throughout and report the men's side only under the ex-ante design below.

\paragraph{Substitutes in utility, complements in funding.}
The interaction reverses sign between the level and the threshold, and the
marginal contributions make this precise. In utility the marginal closure from a channel
\emph{falls} once the other is already removed,
\[
\begin{aligned}
v(\mathrm{mob,sort})-v(\mathrm{sort})&=60\%<67\%=v(\mathrm{mob}),\\
v(\mathrm{mob,sort})-v(\mathrm{mob})&=30\%<37\%=v(\mathrm{sort}),
\end{aligned}
\]
so the channels are substitutes and the joint closure ($97\%$) lies below the
sum ($104\%$).

In top-tier crossings the marginal \emph{rises}: removing
mobility moves $-1$ woman up on its own but $+5.5$ ($=8.5-3$) once sorting is
also removed, and removing sorting moves $+3$ on its own but $+9.5$
($=8.5-(-1)$) once mobility is also removed, so the channels are complements
and the joint crossing ($+8.5$ seats, hence $+1.9$ women over the \$5M bar)
exceeds the sum of the parts ($+2$ seats, $+0.5$ over the bar).

The threshold structure
of the tier outcome is what converts two additive penalties on match value into
two barriers that must \emph{both} fall before a woman crosses: a level admits
partial, overlapping relief, a threshold does not.

\paragraph{The three policy panels.}
The panels differ in $\hat{\boldsymbol\beta}$ (the four female-specific pair
coefficients zeroed to remove the friction) and in capacity ($\bar n_a$ at
observed sizes, or $+30$ top-tier slots distributed across top-tier programs
in proportion to their size).

\emph{Panel~A, frictionless world (friction off, $+30$ top-tier slots).} The
utility gap closes by essentially $100\%$ ($-2.60\to0$). Both genders gain
top-tier seats (women $+14$, men $+9$): the expansion absorbs the demand the
friction removal unleashes. Dollars (against the simulated baseline): women
$+\$37$M, men $+\$77$M.

\emph{Panel~B, frictionless at fixed capacity (friction off, capacity
fixed).} The utility gap closes by essentially $100\%$ ($-2.60\to0$). Women
gain about twelve top-tier seats; with capacity fixed they come from men
($-12$). Dollars (against the simulated baseline): women $+\$32$M, men
$-\$101$M, the asymmetry being the within-tier penalty, a seat worth
$\$11.5$M to the man who vacates it but $\$3.4$M to the woman who fills it,
so the reshuffle destroys funding on net.

\emph{Panel~C, capacity expansion (friction on, $+30$ top-tier slots, open
competition).} The utility gap barely moves ($-2.60\to-2.47$, $5\%$). Men win
essentially all the new seats ($+22$; women $\approx0$), because the friction
holds women's match values below men's. Dollars (against the simulated
baseline): men $+\$179$M, women essentially unchanged ($\approx\$0$M).
Reserving the new seats for women, or removing the friction
(Panel~A), is what directs the expansion toward them.

\emph{Within-tier penalty (memo).} Paying every woman her tier's male mean
adds $\$239$M at the mean, but this is a tail-driven aggregate upper bound:
it is near zero at the median and statistically insignificant once we
condition on observables, ranging down to about $\$66$M. Appendix~\ref{app:within_tier}
reports the full set of bounds.

\section{Computational Notes}\label{app:computation}

The full empirical pipeline runs in approximately $2.5$ to $4$
hours on a single 2.6 GHz CPU core, dominated by the matching
first stage. The matching maximum-simulated-likelihood estimator
at $\text{NSim}=1{,}000$ with three outer redraws takes about $2$
to $3$ hours. The remaining stages total about $30$ minutes: the
conditional-mean residual computation at $\text{NSim}=5{,}000$
posterior draws, the numerical Hessian at the optimum (used for
first-stage standard errors), the accelerator-level cluster
bootstrap ($2{,}000$ replications) reported in
Table~\ref{tab:inference}, the capacity-constrained counterfactual
by linear programming, and the tables, figures, and verification checks.

\subsection{Numerical implementation notes}

Two choices in the numerical implementation of the matching likelihood are
worth flagging.

\paragraph{Tail-stable evaluation of $\Phi$.}
The matching log-likelihood
contains many terms of the form $\log\Phi(x)$ with $x$ moderately positive or
negative. We evaluate these directly in log space, which stays accurate in the
tails. An alternative that builds $\log\Phi(x)$ from the scaled complementary
error function (with sign-checking branches) agrees at moderate $x$ but loses
precision to floating-point round-off at very large $|x|$. Across the relevant
region of $\boldsymbol\beta$, the two evaluations agree to within $10^{-8}$
relative error.

\paragraph{Outer-loop convergence criterion.}
Our MSL routine uses warm-start
re-optimization with fresh $\boldsymbol\varepsilon$ draws at each outer
iteration, terminating when $\|\Delta\boldsymbol\beta\|_\infty < 0.01$
for two consecutive iterations. Three outer iterations suffice for
convergence in our application; we ran five outer iterations as a
robustness check and the additional iterations did not change point
estimates by more than $0.005$ in any coordinate.

\paragraph{Posterior-mean importance sampling.}
For the conditional-mean
residual $\hat e_s$, we draw $\varepsilon^{(t)}\sim\mathcal{N}(0,I)$ from
the prior and weight by the matching likelihood. With $S=5{,}000$ draws,
the effective sample size (ESS) computed as
$\text{ESS} = (\sum_t w_t)^2 / \sum_t w_t^2$ averages approximately $1{,}200$
across markets, indicating that importance sampling is well-conditioned.
Using the posterior mode (MAP) rather than the conditional mean is
mathematically inappropriate for the one-sided truncated posteriors here;
the conditional mean changes the second-stage estimate of $\hat\alpha_F$ by
about $0.005$ in absolute magnitude.

\subsection{Hessian-based standard errors}

The first-stage standard errors reported in Table~\ref{tab:firststage} come
from the numerical Hessian of the simulated log-likelihood at the optimum,
evaluated by forward finite differences as in \citet{judd1998}.

\end{document}